\documentclass[article,nojss,shortnames]{jss}

\usepackage{thumbpdf}
\usepackage{amsfonts,amstext,amsmath,amssymb,amsthm, mathtools}
\usepackage{accents}
\usepackage{rotating}
\usepackage{verbatim}
\usepackage{booktabs}
\usepackage{makecell}
\usepackage{caption}
\usepackage{multirow}
\usepackage{relsize}
\usepackage{diagbox}
\usepackage{numprint}
\usepackage{dcolumn}
\usepackage{pifont}
\usepackage{alltt}
\usepackage{nicefrac}
\usepackage{thmtools, thm-restate}
\usepackage{breqn}
\usepackage{orcidlink}
\usepackage{longtable}

\usepackage[]{graphicx}\usepackage[]{xcolor}
\makeatletter
\def\maxwidth{ \ifdim\Gin@nat@width>\linewidth
    \linewidth
  \else
    \Gin@nat@width
  \fi
}
\makeatother

\definecolor{fgcolor}{rgb}{0.345, 0.345, 0.345}
\newcommand{\hlsng}[1]{\textcolor[rgb]{0.192,0.494,0.8}{#1}}\newcommand{\hldef}[1]{\textcolor[rgb]{0.345,0.345,0.345}{#1}}\newcommand{\hlkwc}[1]{\textcolor[rgb]{0.333,0.667,0.333}{#1}}\newcommand{\hlkwd}[1]{\textcolor[rgb]{0.737,0.353,0.396}{\textbf{#1}}}

\usepackage{framed}
\makeatletter
\newenvironment{kframe}{\def\at@end@of@kframe{}\ifinner\ifhmode \def\at@end@of@kframe{\end{minipage}}\begin{minipage}{\columnwidth}\fi\fi \def\FrameCommand##1{\hskip\@totalleftmargin \hskip-\fboxsep
 \colorbox{shadecolor}{##1}\hskip-\fboxsep
\hskip-\linewidth \hskip-\@totalleftmargin \hskip\columnwidth}\MakeFramed {\advance\hsize-\width
   \@totalleftmargin\z@ \linewidth\hsize
   \@setminipage}}{\par\unskip\endMakeFramed \at@end@of@kframe}
\makeatother

\definecolor{shadecolor}{rgb}{.97, .97, .97}
\definecolor{messagecolor}{rgb}{0, 0, 0}
\definecolor{warningcolor}{rgb}{1, 0, 1}
\definecolor{errorcolor}{rgb}{1, 0, 0}
\newenvironment{knitrout}{}{} 

\usepackage{alltt}

\newcommand{\AUC}{\text{AUC}}
\newcommand{\ROC}{\text{ROC}}

\newcommand{\rY}{Y}

\newcommand{\ry}{y}

\newcommand{\rx}{\xvec}

\newcommand{\h}{h}

\newcommand{\basisy}{\bvec}

\newcommand{\parm}{\varthetavec}
\newcommand{\eparm}{\vartheta}

\newcommand{\eg}{\textit{e.g.}~}

\newcommand{\RR}{\mathbb{R}}

\usepackage{dsfont}

 \DeclareMathOperator{\logit}{logit}

 \DeclareMathOperator{\rank}{rank}

 \DeclareMathOperator{\diag}{diag}

\def \avec {\text{\boldmath$a$}}    \def \mA {\text{\boldmath$A$}}
\def \bvec {\text{\boldmath$b$}}

\def \hvec {\text{\boldmath$h$}}    
    \def \mI {\text{\boldmath$I$}}
    
\def \kvec {\text{\boldmath$k$}}

    \def \mP {\text{\boldmath$P$}}
    \def \mQ {\text{\boldmath$Q$}}
    
\def \svec {\text{\boldmath$s$}}    
    
    \def \mU {\text{\boldmath$U$}}
    
\def \wvec {\text{\boldmath$w$}}    \def \mW {\text{\boldmath$W$}}
\def \xvec {\text{\boldmath$x$}}    \def \mX {\text{\boldmath$X$}}
\def \yvec {\text{\boldmath$y$}}    \def \mY {\text{\boldmath$Y$}}
    \def \mZ {\text{\boldmath$Z$}}

\def \hhat {\text{$\hat h$}}

\def \betavec         {\text{\boldmath$\beta$}}

\def \deltavec        {\text{\boldmath$\delta$}}

\def \thetavec        {\text{\boldmath$\theta$}}
\def \varthetavec     {\text{\boldmath$\vartheta$}}

\def \lambdavec       {\text{\boldmath$\lambda$}}
\def \muvec           {\text{\boldmath$\mu$}}
\def \nuvec           {\text{\boldmath$\nu$}}

\def \deltahatvec        {\text{\boldmath$\hat \delta$}}

\def \mGamma   {\boldsymbol{\Gamma}}

\def \mLambda  {\boldsymbol{\Lambda}}

\def \mSigma   {\boldsymbol{\Sigma}}

\def \nullvec {\mathbf{0}}
\def \onevec {\mathbf{1}}

\declaretheorem[name=Corollary]{corollary}
\declaretheorem[name=Lemma]{lemma}
\declaretheorem[name=Definition,style=definition,numbered=no]{definition}

\author{Ainesh Sewak~\orcidlink{0000-0003-1858-9987} \\ Universit\"at Bern
        \And Sandra Siegfried~\orcidlink{0000-0002-7312-1001} \\ Universit\"at Z\"urich
	\And Torsten Hothorn~\orcidlink{0000-0001-8301-0471} \\ Universit\"at Z\"urich}
\Plainauthor{Ainesh Sewak, Sandra Siegfried, Torsten Hothorn}

\title{Transformation Discriminant Analysis for Constructing Optimal Biomarker Combinations}
\Shorttitle{Optimal Biomarker Combinations}
\Plaintitle{Optimal Biomarker Combinations with Transformation Discriminant Analysis}

\Abstract{
	Accurate diagnostic tests are essential for effective screening and treatment. However, individual biomarkers often fail to provide sufficient diagnostic accuracy, as they typically capture only one aspect of the complex disease process. Combining multiple biomarkers, each capturing a distinct mechanism, can help constructing more informative diagnostic tests. In practice, logistic regression is used as the default to combine biomarkers, but it can perform poorly when biomarker distributions exhibit skewness or differ across disease groups. Nonparametric methods provide more flexibility but generally require large sample sizes that are infrequently available in biomedical research.
We propose a novel framework called transformation discriminant analysis which combines biomarkers through the likelihood ratio function to construct theoretically optimal diagnostic scores. Transformation discriminant analysis balances between flexibility and efficiency. It can accommodate a wide range of distributional shapes and disease-specific dependence structures while remaining fully parametric. This allows for likelihood inference and strong performance even in small-sample settings.
We evaluate TDA through simulations and benchmark its performance against commonly used methods. Finally, we illustrate its utility in constructing an optimal diagnostic test for hepatocellular carcinoma, a disease with no single ideal biomarker. An open-source \proglang{R} implementation is provided for reproducibility and broader application.
 }

\Plainkeywords{diagnostic tests, multivariate transformation model, optimal combination, hepatocellular carcinoma, biomarkers, classification, ROC, AUC}

\Address{
	Ainesh Sewak \\
	Department of Clinical Research \\
	Universit\"at Bern \\
	Freiburgstrasse 3, CH-3010 Bern, Switzerland\\
	Email: \texttt{Ainesh.Sewak@unibe.ch} \\

	Sandra Siegfried, Torsten Hothorn\\
	Institut f\"ur Epidemiologie, Biostatistik und Pr\"avention \\
	Universit\"at Z\"urich \\
	Hirschengraben 84, CH-8001 Z\"urich, Switzerland \\
	Email: \texttt{siegfried.sandra@protonmail.com}, \texttt{Torsten.Hothorn@R-project.org}
}

\begin{document}

\section{Introduction}

Diagnostic testing is central to modern healthcare because it enables the timely identification and management of diseases. Most diagnostic tests rely on individual biomarkers for screening and subsequent treatment decisions~\citep{pepe2005evaluating}. However, the biological heterogeneity of many diseases means that single biomarkers often fail to capture the full picture. This has driven a shift in precision medicine towards using biomarker panels, where multiple markers used together can offer a more comprehensive view of disease pathology~\citep{hartl2023quantitative}. Statistically combining these markers effectively is key to improving diagnostic accuracy.

What is the best way to combine information from multiple biomarkers to discriminate diseased from nondiseased populations? Methodological research has largely focused on linear combinations. Early approaches assumed multivariate normality, leading to solutions like discriminant analysis~\citep{su1993linear}. Later methods relaxed this assumption and optimized empirically based on performance metrics~\citep{pepe2000combining, liu2011min, yin2014optimal, kang2016linear}. However, linear combinations are not necessarily optimal. They may struggle to capture interactions between biomarkers or distributional differences between disease populations~\citep{fong2016combining}.

Theoretically, the path to optimality is clear: the likelihood ratio function provides the uniformly most powerful decision rule for binary classification~\citep{green1966signal, egan1975signal}. Further, any monotonic transformation, such as a risk score, retains this optimality~\citep{mcintosh2002combining}. This has led to logistic regression becoming a default method in practice. However, when data are skewed or exhibit disease-specific dependence structures, logistic regression can yield biased estimates and suboptimal performance~\citep{yan2018combining}. While the likelihood ratio remains to be a viable solution, a gap remains between this theoretical ideal and its practical implementation. Estimating multivariate distributions required for the likelihood ratio is mathematically and computationally challenging~\citep{huang2022linear}.

We address this gap by introducing Transformation Discriminant Analysis (TDA). We propose a flexible multivariate transformation model that estimates the joint distributions of biomarkers and enables the construction of optimal composite diagnostic scores via the likelihood ratio function. TDA generalizes LDA by operating on a transformed scale and accommodates key clinical complexities such as skewed marginals, disease-specific dependence structures and missing biomarkers. Its parameterization supports efficient computation of diagnostic metrics such as ROC curves and AUC, along with standard likelihood-based inference. We further develop model assessment techniques to evaluate goodness-of-fit of the underlying model.

We demonstrate TDA’s utility in the context of hepatocellular carcinoma (HCC), the most common form of liver cancer. Current diagnostic practices for HCC rely on imaging, biopsy, and serum biomarkers. Among these, alpha-fetoprotein (AFP) is most commonly used, but its standalone diagnostic performance is limited, especially in patients with benign liver conditions~\citep{di2005serum}. To address this limitation, alternative biomarkers have been proposed~\citep{de2018novel}. Using data from a retrospective case-control study~\citep{jang2016diagnostic}, we construct and evaluate an optimal diagnostic score that combines AFP with additional biomarkers to improve HCC detection.

In the sections that follow, we first introduce the notation and briefly review existing methods for combining multiple biomarkers. We then present the TDA framework, describing the multivariate transformation model, the associated likelihood ratio function, special cases, connections to existing models and estimation procedures. The method’s performance is evaluated through simulation studies, followed by an application to hepatocellular carcinoma diagnosis. We discuss the results and outline future directions. We conclude by introducing an \proglang{R} add-on package that implements TDA and provides reproducibility materials.

\section{Optimal biomarker combinations}
\label{sec:opt}

\subsection{Notation}

We focus on data derived from case-control study designs, where $D$
represents a binary random variable indicating the absence ($D = 0$, denoting a
nondiseased subject) or presence ($D = 1$) of a specific disease, such as
histologically confirmed HCC from our application. The
random vector $\mY_d = \{\mY \mid D = d\} = (Y_{d1}$, $Y_{d2},\dots,
Y_{dJ})^\top \in \RR^J$ represents the $J$ absolutely continuous biomarker
observations of a subject with disease status $D=d$.  Let $f_d : \RR^J \mapsto
\RR^+$ be the absolutely continuous joint conditional probability density
function (PDF) of the biomarkers, with $f_0$ characterizing the biomarker PDF
for the nondiseased population and $f_1$ for the diseased population. The
development of an optimal diagnostic score is based on data obtained from
independent observations, denoted as $i=1, \dots, N = N_0 + N_1$, originating
from both nondiseased and diseased populations.

The primary objective of this paper is to derive a scalar-valued
function, denoted by $L : \RR^J \mapsto \RR$. This function combines the $J$
biomarkers into a composite diagnostic score and we aim at finding a function $L$
maximizing diagnostic accuracy. The composite score can be employed to classify a yet undiagnosed subject
based
on observed biomarker values $\yvec = (\ry_1, \dots, \ry_J) \in \RR^J$.
The classification involves designating a subject as diseased when $L(\yvec) > c$ for a specified threshold $c \in \RR$. 

Define $L_0 = \log(L(\mY_0)) \sim G_0$ and $L_1 =
\log(L(\mY_1)) \sim G_1$ as the log-transformed random variables of the resulting composite scores
in the nondiseased and diseased populations, with $G_0$ and $G_1$ being their
respective absolute continuous cumulative distribution functions (CDFs).
Specificity and
sensitivity, representing the probability of accurately identifying nondiseased
or diseased subjects, are defined by $\Prob(L_0 \leq c) = G_0(c)$ and
$\Prob(L_1 > c) = 1 - G_1(c)$. The ROC curve graphically summarizes the
trade-off between sensitivity and specificity, serving as a quantifiable measure
for diagnostic accuracy in an optimal test. The 
optimal ROC curve, based on the composite score, is given by $\ROC(p) = 1 - G_1
\left( G_0^{-1}(1- p) \right)$.

\subsection{Related work}

Various approaches have been proposed for developing combinations of multiple biomarkers for discriminating between two populations. Linear combinations of biomarkers aim to find the best set of coefficients $\avec \in \RR^J$ such that the composite diagnostic scores $L(\mY_d) = \avec^\top \mY_d$ maximize discrimination between disease populations. 
Under the assumption of multivariate normality for $\mY_d$, the classical linear discriminant analysis (LDA) coefficient yields the optimal linear combination~\citep{su1993linear}. To avoid distributional assumptions, later methods focused on optimizing empirical performance criteria such as the AUC~\citep{pepe2000combining, pepe2006combining, huang2022linear}, partial AUC~\citep{hsu2014biomarker, yan2018combining}, or the Youden index~\citep{yin2014optimal}. These methods are flexible and provide distribution-free linear combinations of biomarkers. However, linearity imposes some limitations. Such scores may fail to capture skewed biomarker distributions, interactions or nonlinearities that often arise in biomedical applications~\citep{fong2016combining}.

From a theoretical perspective, the likelihood ratio function
\begin{align*}
	L(\mY_d) = \frac{f_1(\mY_d)}{f_0(\mY_d)},
\end{align*}
provides the optimal combination under the Neyman-Pearson lemma~\citep{mcintosh2002combining}. When used as a composite diagnostic score, the resulting ROC curve maximizes sensitivity at every level of specificity~\citep{pepe2003statistical}. All associated performance criteria such as the AUC, pAUC and Youden Index are also maximized. By Bayes’ theorem, the likelihood ratio is also a monotone function of the posterior odds of disease
\begin{align}
	\log\left(\frac{\Prob(D = 1 \mid \mY = \yvec)}{\Prob(D = 0 \mid \mY = \yvec)}\right) = \log\left(\frac{\Prob(D = 1)}{\Prob(D = 0)}\right) + \log\left(\frac{f_1(\yvec)}{f_0(\yvec)}\right).
\end{align}
This decomposition highlights two additional modeling strategies. 
\emph{Discriminative} methods, such as logistic regression, model the left-hand side, the log-odds of disease, often assuming a linear form. Flexible machine learning techniques can also be used to estimate $\Prob(D = 1 \mid \mY)$ directly~\citep{pepe2006combining}, though they often require large sample sizes and lack interpretability.

\emph{Generative} methods, in contrast, model the disease-specific densities $f_0$ and $f_1$. Despite its name, LDA is a generative method. When correctly specified, such models can be more statistically efficient. For example, the logistic regression is only two-thirds as efficient as LDA under multivariate normality~\citep{efron1975efficiency} and it is advocated against the default use of logistic regression due to its inefficiency in nonnormal settings~\citep{oniell1980general}. Still, logistic regression remains widely used due to its simplicity, robustness across many settings (e.g., exponential families) and interpretability~\citep{press1978choosing}. However, it cannot accommodate nonnormal settings or disease-specific dependence structures unless explicitly extended~\citep{kay1987transformations}.

Apart from the multivariate normal case, flexibly estimating the full joint densities $f_0$ and $f_1$ is challenging in practice. Some authors have proposed modeling the likelihood ratio directly~\citep{qin2010best, chen2016using, martinez2021optimal} to circumvent this difficulty. Others use generative semiparametric approaches, but these often rely on tuning parameters or are difficult to estimate at scale.

In this paper, we propose a generative method called transformation discriminant analysis (TDA). This is a fully parametric multivariate transformation model which can flexibly estimate the disease-specific densities and thus the likelihood ratio function. This allows us to construct composite diagnostic scores that achieve the optimal ROC curve under the Neyman–Pearson framework. Our model captures varying biomarker distribution shapes and disease-specific dependence structures. Its generative formulation naturally accommodates missing biomarkers and detection limits. Unlike existing generative transformation-based methods~\citep{Lafferty_Liu_Wasserman_2012, lyu2019new, kim2015semiparametric, du2024likelihood}, our approach is not sensitive to tuning parameters and enables likelihood-based inference including confidence intervals for model parameters and diagnostic accuracy metrics such as the AUC. Particularly in small-sample biomedical applications, this could lead to asymptotic efficiency without sacrificing theoretical optimality.

\section{Multivariate transformation model}
\label{sec:methods}

We propose a multivariate
transformation model featuring an unknown 
transformation function $\hvec_d: \mathbb{R}^J \mapsto \mathbb{R}^J$ to model the joint density and account for the correlation between biomarkers~\citep{klein2022multivariate}.  This
function is defined coordinate-wise on the observed biomarkers, $\hvec_d(\yvec) =
(h_{d1}(y_1), \dots, h_{dJ}(y_J))^\top$ and is 
monotonically nondecreasing in each of its coordinates.

The purpose of this transformation is to map the unknown distribution of $\mY_d$
for a given disease indicator $D=d \in \{0, 1\}$ to a random vector with a known
distribution, denoted as $\mZ_d = \hvec_d(\mY_d)$. Specifically, the vector
$\mZ_d = (Z_{d1}, \dots, Z_{dJ})^\top$ follows a zero-mean multivariate normal
distribution, $\mZ_d \sim N_J(\nullvec, \mSigma_d)$, with a disease-dependent
covariance matrix $\mSigma_d \in \mathbb{R}^{J \times J}$. The entries of the
covariance matrix measure the dependence between the \emph{transformed}
biomarkers in each of the populations. The joint CDF of $\mY_d$ is given by
\begin{align*}
	\Prob(\mY \leq \yvec \mid D=d) = \Prob(\mY_d \le \yvec)&= \Prob \left(\hvec_d(\mY_d) \leq \hvec_d(\yvec) \right)\\
	& = \Prob(\mZ_d \leq \hvec_d(\yvec) ) = {\Phi}_{\nullvec, \mSigma_d} \left(h_{d1}(y_1), \dots, h_{dJ}(y_J)  \right)
\end{align*}
where {\large $\Phi_{\nullvec, \mSigma}$} is the joint CDF of a multivariate
normal distribution with a zero mean vector and covariance matrix $\mSigma$.
To ensure identifiability, we standardize this matrix such that $\diag(\mSigma_d) =
\onevec$ and $\mSigma_d$ is a correlation matrix. This leads to the
interpretation of $h_{dj}$ as marginal distribution functions on the probit
scale $\Prob(\rY_{j} \le \ry_j \mid D = d) = \Prob(\rY_{dj} \le \ry_j) = \Phi({h}_{dj}(y_{j}))$. The following proposition provides the optimal function for combining multiple
biomarkers under the multivariate transformation model.
\begin{restatable}{proposition}{TDA}
Suppose the derivatives of the marginal transformation functions exist such that
$h'_{dj}(y_j) >0$ for $j=1, \dots, J$ and let the joint PDF of the biomarkers be
\begin{align*}
	f_d(\yvec) = {\phi}_{\nullvec, \mSigma_d} \left(h_{d1}(y_1), \dots, h_{dJ}(y_J) \right) \prod_{j=1}^{J} h'_{dj}(y_j),
\end{align*}
where {\large $\phi_{\nullvec, \mSigma}$} is the joint PDF of a multivariate
normal distribution with a zero mean vector and correlation matrix $\mSigma$.
Then the log-likelihood ratio function is
\begin{align*}
	\log(L(\yvec)) = -\frac{1}{2}\left(  \log \left( \frac{\lvert \mSigma_1 \rvert}{\lvert \mSigma_0 \rvert}\right) + \hvec_1(\yvec)^\top \mSigma_1^{-1} \hvec_1(\yvec) - \hvec_0(\yvec)^\top \mSigma_0^{-1} \hvec_0(\yvec)\right) + \sum_{j=1}^J \log \left(\frac{h'_{1j}(y_j)}{h'_{0j}(y_j)}\right),
\end{align*}
where $\lvert \mSigma_d \rvert \neq 0$ is the determinant of the matrix
$\mSigma_d$.
\label{prop:TDA}
\end{restatable}
The proof of Proposition~\ref{prop:TDA} is given in Appendix~\ref{sec:proof} and follows from the definition of the likelihood
ratio function. We call all ROC curves and AUCs derived from the likelihood ratio function under the multivariate transformation model as \emph{model-based} ROC curves and AUCs.

\subsection{Location-scale marginal model}
The general multivariate transformation model, as presented above, incorporates
fully flexible marginal distributions for the biomarkers in each of the
nondiseased and diseased classes. This model requires simulations for the
sampling distributions of $L_0$ and $L_1$, particularly when calculating the
optimal model-based ROC curve and corresponding AUC, given the absence of simple closed-form
expressions. We give a location-scale simplification of the marginal model in
the following proposition which ensures analytical accessibility to these
distributions, while requiring fewer overall model parameters.

\begin{restatable}{proposition}{lsTDA}
Assume a common transformation function $\hvec: \mathbb{R}^J \mapsto
\mathbb{R}^J$ with $\hvec(\yvec) = (h_1(y_1), \dots, h_J(y_J))^\top$ such that
the $j$th marginal transformation function is defined as
\begin{align*}
	h_{dj}(y_j) = \frac{h_j(y_j) - \delta_j d}{\exp(\gamma_j d)} \quad \text{for} \; j = 1,\dots, J,
\end{align*}
where $\delta_j \in \RR$ and $\exp(\gamma_j) \in \RR^+$. Then the multivariate
model can be expressed as
\begin{align*}
	\hvec(\mY_d) = \deltavec_d + \mGamma_d^{-1} \mZ_d,
\end{align*}
where $\deltavec_0 = \nullvec$, $\deltavec_1 = \deltavec = (\delta_1, \dots, \delta_J)^\top$, $\mGamma_0^{-1} = \mI$, $\mGamma_1^{-1} = \mGamma^{-1} = \diag(\exp(\gamma_1), \dots, \exp(\gamma_J))$, $\mZ_d \sim N_J(\nullvec, \mSigma_d)$ and the log-likelihood ratio function is
	\begin{align*}
		\log(L(\yvec)) \propto (\hvec(\yvec) - \betavec)^\top \mA (\hvec(\yvec) - \betavec),
	\end{align*}
	\sloppy
	where $\mA = \mGamma \mSigma_1^{-1} \mGamma - \mSigma_0^{-1}$ and $\betavec = \left( \mI + (\mSigma_0 \mA)^{-1} \right) \deltavec$.
\label{prop:lsTDA}
\end{restatable}
The proof of Proposition~\ref{prop:lsTDA} is given in Appendix~\ref{sec:proof} and follows from Proposition~\ref{prop:TDA}. The
parameters within the marginal models are also interpretable as follows. The
location term $\delta_j$ accommodates distinct baseline biomarker levels for
diseased and nondiseased cases while $\exp(\gamma_j)$ represents the scaling
term allowing for different degrees of dispersion based on disease status
\citep{Siegfried_Kook_Hothorn_2023}.

The following corollary directly arises from Proposition~\ref{prop:lsTDA} and allows for fast computation of diagnostic accuracy metrics for the composite score $L_d$, including model-based ROC curves and AUCs. This approach involves evaluating a univariate generalized chi-square distribution, defined as a weighted sum of non-central chi-square distributions~\citep{cacoullos1984quadratic}, whose parameters are derived from the coefficients of the location-scale multivariate model.
\begin{restatable}{corollary}{gchisq}
Let the spectral decomposition of $\tilde{\mSigma}_d^{\frac{1}{2}} \mA
\tilde{\mSigma}_d^{\frac{1}{2}}$ be given by $\mP_d \mW_d \mP_d^\top$ where $\tilde{\mSigma}_d$ is the correlation matrix of $\hvec(\mY_d)$. Then the scalar composite score $L_d$ follows a generalized chi-square distribution $G_{\chi_J^2}(\wvec_d, \nuvec_d)$ with weights as $\wvec_d = \diag(\mW_d)$, the non-centrality parameters
$\nuvec_d = \diag(\mP_d^\top \mSigma_d^{-\frac{1}{2}}(\deltavec_d -
\betavec))^2$ and the degrees of freedom $\onevec \in \RR^J$.
\label{cor:gchisq}
\end{restatable}
Note that $\mP_d$ is an orthogonal matrix whose columns are the real,
orthonormal eigenvectors and $\mW_d$ is a diagonal matrix whose entries are the
eigenvalues. The proof for this result can be found in Appendix~\ref{sec:proof}. It follows from the fact that the log-likelihood
ratio function under this model takes a quadratic form and thus the
resulting distribution of the scalar composite score $L_d$ is also in a
quadratic form of a multivariate normal variable.

The generalized chi-square distribution lacks a closed-form distribution
function, but efficient computational methods have been developed for its
evaluation \citep{davies1980distribution}. Therefore, as the optimal model-based ROC curve
is defined by the distribution functions of $L_0$ and $L_1$, it can be
calculated directly from the model parameters, along with the corresponding AUC.

\subsection{Relationship to LDA}

In a specific case of our model, we arrive at the same result as LDA, which is the established as the optimal linear combination of
biomarkers~\citep{su1993linear}. However, our approach extends this result to include combinations of
transformed biomarkers without imposing the assumption of normality on $\mY_d$.
\begin{corollary}
Assume a global covariance matrix $\mSigma_d = \mSigma$ for both classes,
with a single multivariate transformation function $\hvec_d = \hvec$, and
omitting scaling terms $\mGamma_d = \mI$ for $d=0,1$. Then the log-likelihood ratio
function is $\deltavec^\top \mSigma^{-1} \left(\hvec(\yvec) - \frac{1}{2} \deltavec\right)$.
\label{cor:lTDA}
\end{corollary}
Given that the log-likelihood ratio is linear in the vector of
transformed biomarkers $\hvec(\yvec)$, the best linear combination of the
transformed biomarkers is proportional to the Fisher's discriminant coefficient
$\deltavec^\top \mSigma^{-1}$. In this model, the distribution of the transformed biomarkers in the nondiseased
class is given by $\hvec(\mY_0) \sim N_J(\nullvec, \mSigma)$, and in the
diseased class, it is $\hvec(\mY_1) \sim N_J(\deltavec, \mSigma)$. Consequently,
the distributions of the composite scores are $L_0 \sim N\left(-\frac{1}{2}
\deltavec^\top \mSigma^{-1} \deltavec, \deltavec^\top \mSigma^{-1}
\deltavec\right)$ and $L_1 \sim N\left(\frac{1}{2} \deltavec^\top \mSigma^{-1}
\deltavec, \deltavec^\top \mSigma^{-1} \deltavec\right)$. The optimal model-based ROC curve
is binormal and can be expressed as
\begin{align} \label{fm:ROC}
	\ROC(p) = 1 - \Phi \left( \Phi^{-1}(1 - p) - \sqrt{\deltavec^\top \mSigma^{-1} \deltavec} \right),
\end{align}
while the AUC is given by
\begin{align} \label{fm:AUC}
	\AUC = \Phi \left( \sqrt{\frac{\deltavec^\top \mSigma^{-1} \deltavec}{2}} \right).
\end{align}
In practice, and also in our application, the differences between the biomarker
distributions of diseased and nondiseased cases often extend beyond linear
location shifts, even on the transformed scale defined by $\hvec$. This
necessitates additional considerations, such as scaling, for appropriate
modeling.

Because of this connection of the log-likelihood ratio function derived
from the multivariate transformation model with Fisher's discriminant analysis, in the following we
refer to disease classifiers based on $\log(L(\yvec))$ as \emph{transformation
discriminant analysis} (TDA).

\subsection{Parameterization and estimation}

We parameterize the transformation function $\hvec_d(\yvec) = \hvec_d(\yvec \mid
\thetavec_d)$ via parameters $\thetavec_d = (\parm_{d1}^\top, \dots,
\parm_{dJ}^\top)^\top \in \RR^{J(M+1)}$. The $j$th element $\parm_{dj}$ of this
vector parameterizes the marginal transformation function $\h_{dj}(y) =
\h_{dj}(y \mid
\parm_{dj})$ for the $j$th biomarker in either diseased ($d = 1$) and nondiseased
($d = 0$) classes. We describe the $j$th marginal
transformation function in terms of a monotonically nondecreasing polynomial in
Bernstein form
\begin{align}
	\h_{dj}(y \mid \parm_{dj}) = \basisy_j(y)^\top \parm_{dj} = \sum_{m=0}^{M} \eparm_{djm} b_{jm}(y) \quad \text{for} \quad y \in \RR, 
\end{align}
where~$\basisy_j(y)=(b_{j0}(y),\dots,b_{jM}(y))^\top$ is a vector of $M+1$ basis
functions with associated coefficients~$\parm_{dj} = (\eparm_{dj1}, \dots,
\eparm_{dj(M+1)})^\top \in \RR^{M+1}$ for $j=1,\dots,J$~\citep{hothorn2018most, klein2022multivariate}. The Bernstein basis
polynomial of order~$M$ is defined on the interval~$[l,u]$ as
\begin{align}
	\label{eq:basis}
	b_{jm}(y) = {M \choose m} \tilde{y}^m (1-\tilde{y})^{M-m}, \quad m=0,\dots,M,
\end{align}
where $ \tilde{y} = \frac{y-l}{u-l} \in [0,1]$. The constraint $\eparm_{djm}
\leq \eparm_{dj(m+1)}$ for~$m=0,\dots,M-1$, guarantees the monotonicity of
$\h_{dj}$ and the smooth parameterization the existence of the derivative
$\h^\prime_{dj}(y \mid \parm_{dj}) = \basisy^\prime_j(y)^\top \parm_{dj}$.
Weierstrass' theorem ensures that any real-valued continuous function can be
approximated uniformly on the interval $[l, u]$ with increasing order
$M$~\citep{farouki2012bernstein}. With large enough orders $M$, the marginal
distribution functions $\Prob(\rY_{dj} \le \ry_j) =
\Phi(\basisy_j(\ry_j)^\top \parm_{dj})$ closely approximate the marginal
empirical cumulative distribution functions (ECDFs) used in the partially
nonparametric estimation of nonparanormal models
\citep{Lafferty_Liu_Wasserman_2012}. The benefit of compromising marginal fit by
using moderate orders $M$ is the ability to formulate and optimize the
log-likelihood simultaneously for all marginal parameters and the parameters
defining the unstructured Gaussian copula.

We parameterize $\mSigma_d$ defining this copula in terms of the Cholesky factor
of the precision matrix $\mSigma_d^{-1} = \tilde{\mLambda}_d^\top
\tilde{\mLambda}_d$ and standardize the lower triangular $J \times J$ unit
matrix $\mLambda_d$ such that $\tilde{\mLambda}_d =  \mLambda_d
\diag(\mLambda_d^{-1}
\mLambda_d^{-\top})^{\nicefrac{1}{2}}$ and thus $\diag(\mSigma_d) = \onevec$.
From a computational point of view, quadratic forms ${\hvec}_d(\yvec \mid
\thetavec_d)^\top \mSigma_d^{-1} {\hvec}_d(\yvec \mid \thetavec_d)$ and
determinants $\lvert \mSigma_d \rvert$ arising in the log-likelihood and
log-likelihood ratio functions simplify to $ {\hvec}_d(\yvec)^\top
\tilde{\mLambda}_d^\top \tilde{\mLambda}_d {\hvec}_d(\yvec)$ and $\lvert
\mSigma_d \rvert = \left(\prod_{j = 1}^J
\diag(\tilde{\mLambda}_d)_j\right)^{-2}$, respectively. The term
$\sqrt{\deltavec^\top \mSigma^{-1} \deltavec}$ in (\ref{fm:ROC}) and (\ref{fm:AUC})
simplifies to $\lVert \tilde{\mLambda} \deltavec \rVert_2$ when $\mSigma_d =
\mSigma$. 
Furthermore, the lower
triangular elements $\lambdavec_d \in \RR^{\nicefrac{J(J-1)}{2}} $ of
$\mLambda_d$ are unconstrained yet lead to a positive definite correlation
matrix $\mSigma_d(\lambdavec_d)$.

The log-likelihood contribution of a single observation $\yvec$ being an absolutely continuous vector of biomarker
results, is given by
\begin{align*}
	\ell(\thetavec_d, \lambdavec_d \mid \yvec) = \log \left( {\phi}_{\nullvec, \mSigma_d(\lambdavec_d)} \left(
\hvec_d(\yvec \mid \thetavec_d) \right)\right) + \sum_{j=1}^J \log \left( \h'_{dj} \left( y_j \mid \parm_{dj} \right) \right) \quad \text{for} \; d=0,1.
\end{align*}
The maximum likelihood estimate of $(\thetavec_d, \lambdavec_d)$ is derived from
a sample of $N_d$ independent and identically distributed observations from
either diseased ($d = 1$) or nondiseased ($d = 0$) subjects using constrained
maximization algorithms. While the general parameterization and estimation procedure are detailed elsewhere~\citep{klein2022multivariate}, our application introduces a novel standardization of the Cholesky factors $\tilde{\mLambda}_d$.
This standardization ensures that each $j$th transformation function can be interpreted as a marginal distribution function on the probit scale, which is essential for deriving model-based AUCs and ROC curves. Disease-specific score functions and theoretical properties of the likelihood-based inference are described in \citep{hothorn2024on}.

Note that in the case of location-scale marginal models, the transformation
functions $\hvec_d$ share parameters between both classes. Each marginal
transformation function is then
\begin{align}
\h_{dj}(y \mid \parm_{j}, \delta_j, \gamma_j) = \frac{\basisy_j(y)^\top \parm_{j} - \delta_j d}{\exp(\gamma_j d)}
\end{align}
and one has to maximize the joint likelihood of both diseased and undiseased
subjects with respect to the common parameters $\thetavec = \left( (\parm_1,
\delta_1, \gamma_1)^\top, \dots, (\parm_J, \delta_J, \gamma_J)^\top)\right)^\top
\in \RR^{J(M + 1) + 2J}$ in addition to $\lambdavec_0$ and $\lambdavec_1$ (which
one might want to be equal) based on all $N = N_0 + N_1$ subjects.

\subsection{Alternative marginal distributions}

With $F: \RR \rightarrow [0, 1]$ denoting an absolutely continuous cumulative distribution function
with a log-concave density,  the marginal distributions in our framework can be expressed as $\Prob(\rY_{dj} \le \ry_j) = F(\h_{dj}(\ry_j))$, where 
where $h_{dj}$ is a monotonically increasing transformation function. In the joint multivariate model this leads to the following transformation function
\begin{align*}
	\hvec_d(\yvec \mid \thetavec_d) = \left(
	\Phi^{-1}(F(\h_{d1}(\ry_1 \mid \parm_1))), \dots, \Phi^{-1}(F(\h_{dJ}(\ry_J \mid \parm_J)))\right)^\top.
\end{align*}
These choices are analogous to link functions in generalized linear models (GLMs). For example, using $F = \text{logit}^{-1}$ leads to a log-odds interpretation, while $F = \text{cloglog}^{-1}$ corresponds to hazard-based interpretations. Different selections of $F$ lead to alternative marginal models and influence the interpretation of location and scale parameters. In our application, we use $F=\Phi$ (the standard normal CDF) for computational simplicity, but more robust alternatives such as $F = \text{logit}^{-1}$ may be preferred in practice due to their interpretability and robustness properties \citep{sewak2023estimating}.

\subsection{Missing and censored biomarkers}
\label{sec:subset}
Our proposed multivariate transformation model derives an optimal diagnostic score by combining multiple biomarkers through a likelihood ratio function. However, in practice, biomarker measurements may be partially missing due to feasibility constraints or only a subset may be available at test time. Our framework accommodates such cases by computing likelihood ratios using the marginal distribution of the observed biomarkers.

Without loss of generality, suppose biomarker $Y_1$ is missing at random and we observe the subset $\yvec^* = (y_2, \dots, y_J)$. By modeling the full joint distribution of $\mY_d$ and applying the law of total probability, we obtain the marginal likelihood
\begin{align*}
	f_d(\yvec^*) = {\phi}_{\nullvec, \mSigma_d^{-1, -1}} \left(h_{d2}(y_2), \dots, h_{dJ}(y_J) \right) \prod_{j=2}^{J} h'_{dj}(y_j),
\end{align*}
where $\mSigma_d^{-1, -1}$ denotes the removal of the first row and column of $\mSigma_d$. This allows us to compute the likelihood ratio
\begin{align*}
	L(\yvec^*) = \frac{f_1(y_2,\dots,y_J)}{f_0(y_2,\dots,y_J)}
\end{align*}
and classify subjects even when certain biomarkers are unobserved. Similarly, under the location-scale simplification, one can compute the model-based ROC curve and AUC using only the relevant subset of parameters $\deltavec_d^{-1}$ and $\mGamma_d^{-1}$.

More generally, the nonparanormal likelihood formulation permits direct incorporation of missingness into the estimation procedure~\citep{hothorn2024on}. Rather than discarding partially observed cases, our method uses the observed biomarker values for each subject, leading to potentially greater efficiency. The same framework extends naturally to biomarkers that are subject to lower or upper limits of detection. Here, we can treat the affected values as censored and then the likelihood integrates over the censored regions. A detailed worked example for time-varying prognostic biomarkers is given in \citep{sewak2025npb}.
 
\section{Empirical evaluation}
\label{sec:empeval}

We assessed the performance of transformation discriminant analysis (TDA) for optimally combining multiple biomarkers for disease diagnosis using simulation studies. The aims were to: (i) evaluate performance under commonly assumed data generating processes; and (ii) assess the effects of model misspecification.

\subsection{Methods compared}

We compared six variants of the proposed multivariate transformation model discussed in Section~\ref{sec:methods}: disease-specific marginal transformations (sTDA), location-scale marginal transformations (lsTDA) and location-only marginal transformations (TDA), each with either a global or disease-specific correlation structure. The latter are denoted as sTDA$_d$, lsTDA$_d$, and TDA$_d$.

As benchmarks, we included popular classification methods: logistic regression (LR), generalized additive models (GAM), random forests (RF) and eXtreme Gradient Boosting (XGBoost). We also evaluated classical discriminant analysis approaches: linear (LDA), quadratic (QDA), mixture (MDA), and flexible discriminant analysis (FDA). Additionally, we implemented linear combination methods proposed in the biomarker literature, including the step-down approach~\citep{kang2016linear} (KT) and the min-max method~\citep{liu2011min} (LIU). For reference, we also computed the AUC corresponding to the true likelihood ratio.

\subsection{Simulation setup and scenarios}

We considered four biomarkers to match the HCC application in Section~\ref{sec:application} and simulated data under five distinct scenarios. Each scenario was evaluated at sample sizes $N \in \{50, 100, 200\}$, with equal numbers of diseased and nondiseased subjects. These sample sizes reflect those commonly encountered in medical studies focused on developing or validating biomarker combinations. 
In prior simulation studies, prevalence had minimal impact on most methods and was thus held constant in our study at 50\%~\citep{du2024likelihood}. We generated 1{,}000 replications per scenario, with an independent large test dataset of size 10{,}000 for out-of-sample evaluation. Performance was assessed using out-of-sample (OOS) AUC. Additional metrics, including mean squared error (MSE) of the AUC are reported in the Appendix.

\subsubsection{Scenario A: Multivariate normal biomarkers}

This scenario represents the ideal setting for classical LDA and logistic regression. We generated data from multivariate normal distributions with equal covariance matrices for diseased and nondiseased groups. 
Means were $\muvec_0 = (0, 0, 0, 0)^\top$ for nondiseased and $\muvec_1 = (-0.2, 0.3, 0.7, -0.1)^\top$ for diseased subjects. These shifts were chosen to yield an approximate true AUC of 0.8. The common covariance matrix $\mSigma = \mSigma_0 = \mSigma_1$ was estimated from the HCC biomarker data
\[
\mSigma = 
\begin{pmatrix}
	1.00 & 0.17 & 0.36 & 0.32 \\
	0.17 & 1.00 & 0.41 & 0.45 \\
	0.36 & 0.41 & 1.00 & 0.82 \\
	0.32 & 0.45 & 0.82 & 1.00 \\
\end{pmatrix}
\]

\subsubsection{Scenario B: Multivariate skewed biomarkers}

To evaluate robustness to skewed distributions, we considered two variants.
\begin{itemize}
	\item[(i)] Multivariate log-normal distributions with log-scale means matched to Scenario A.
	\item[(ii)] Biomarkers generated from the following skewed distributions:
	\begin{itemize}
		\item Nondiseased: $(N(0.6, 1),\; \chi^2(2.5),\; \mathrm{Exp}(1),\; \Gamma(1.2, 1))$
		\item Diseased: $(N(1.1, 1),\; \chi^2(3),\; \mathrm{Exp}(1.7),\; \Gamma(2, 1))$
	\end{itemize}
\end{itemize}
Here, $\chi^2(k)$ represents the chi-squared distribution with $k$ degrees of freedom, $\text{Exp}(\lambda)$ is the exponential distribution with rate $\lambda$, and $\Gamma(\alpha, \beta)$ represents the gamma distribution with shape parameter $\alpha$ and rate parameter $\beta$. Covariance matrices were the same as in Scenario~A.

\subsubsection{Scenario C: Disease-specific dependence}

To reflect disease-specific dependence structures between biomarkers, we allowed different correlation matrices in the two disease groups. Marginals were kept identical to Scenarios A or B.

Correlation matrices were estimated from the HCC dataset:
\[
\mSigma_0 =
\begin{pmatrix}
	1.00 & 0.05 & 0.24 & 0.10 \\
	0.05 & 1.00 & 0.23 & 0.35 \\
	0.24 & 0.23 & 1.00 & 0.62 \\
	0.10 & 0.35 & 0.62 & 1.00 \\
\end{pmatrix}
\quad
\mSigma_1 =
\begin{pmatrix}
	1.00 & 0.17 & 0.33 & 0.31 \\
	0.17 & 1.00 & 0.41 & 0.40 \\
	0.33 & 0.41 & 1.00 & 0.92 \\
	0.31 & 0.40 & 0.92 & 1.00 \\
\end{pmatrix}
\]

Such differences in structure are known to reduce the performance of methods assuming homogeneous dependence such as logistic regression~\citep{yan2018combining}.

\subsubsection{Scenario D: Tail dependence}

To investigate the impact of dependence misspecification, we generated data from the Clayton (strong lower tail dependence) and Gumbel copulas (upper tail dependence). Copula parameters were estimated from the HCC application dataset (Clayton: $\theta=0.4146$; Gumbel: $\theta=1.3170$).

\subsubsection{Scenario E: Logistic model}

In this setting, data were generated directly from logistic regression models, for which discriminative approaches are well suited and the TDA modeling assumptions are violated. We evaluated two settings:
\begin{itemize}
	\item Linear model: log odds of disease as a linear function of $\mY$. i.e., 
	$\logit(\Prob(D = 1 \mid \mY) = \beta_0 + \betavec^\top \mY)$, where $\mY \sim N(0, \onevec_4)$, $\beta_0 = 0.5$ and $\betavec = \left(0.5, -0.6, 1.1, 0.4 \right)$ was chosen to yield a true AUC of approximately 0.80.
	\item Interaction model: log odds of disease including pairwise interactions between biomarkers.
	\begin{align*}
		\logit(\Prob(D = 1 \mid \mY)) = -0.5 + 1.2 Y_1 - 0.8 Y_2 + 0.6 Y_3^2 - 0.4 Y_4^2 + 0.7 Y_1 Y_2 - 0.5 Y_3 Y_4.
	\end{align*}
\end{itemize}
In both cases, the class label $D$ was generated via Bernoulli sampling from the implied probability. This scenario tests the robustness to model misspecification for generative methods like TDA when the true model aligns with a discriminative paradigm.

\subsection{Results}

\begin{figure}[t!]
	\centering 
	\includegraphics[width=\linewidth]{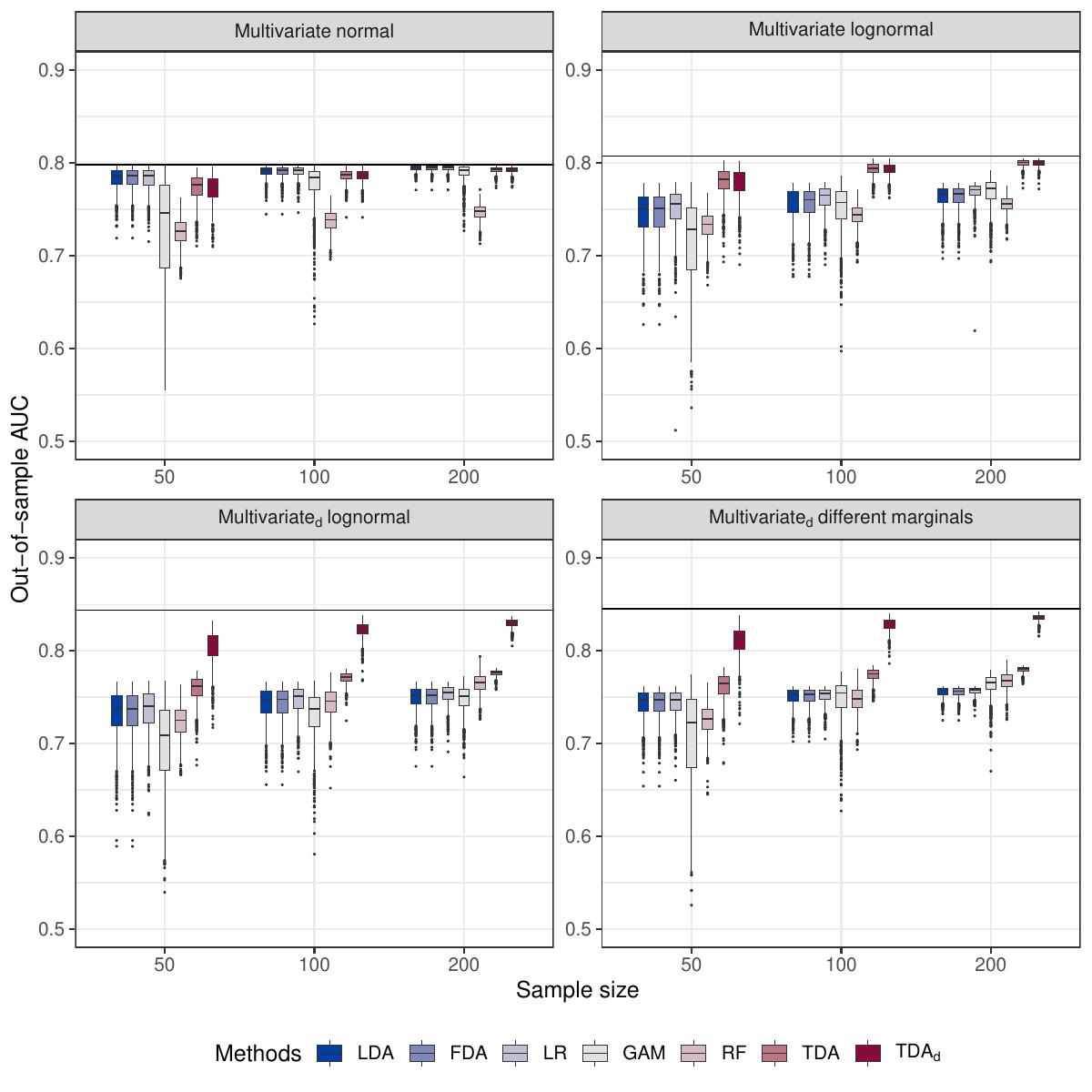} 
	\caption{Empirical out-of-sample area under the receiver operating
	characteristic curve (AUC) across different simulation settings (Scenarios A, B and C). Box-plots are color-coded to categorize methods, with our proposed
	approaches represented by TDA. The optimal AUC in each setting is marked by the black line.}
	\label{fig:aucs}
\end{figure}

We summarized the results of the empirical out-of-sample AUCs under Scenario A, B and C in Figure~\ref{fig:aucs}. For the main text, we present results for the lTDA and lTDA$_d$ models, which we consider as default parameterizations for simplicity and performance. For comparison, we use LDA, FDA, LR, GAM and RF as the most commonly applied alternatives. Complete results for all methods are available in the Appendix Table~\ref{tab:sim}.

\subsubsection{Scenarios A-B: Gaussian and skewed marginals}

When biomarkers followed a multivariate normal distribution, all methods performed similarly and approached the optimal AUC with increasing sample sizes. However, under skewed marginals (log-normal or different marginal distributions), performance differed. LR and LDA declined due to violated normality assumptions. GAM showed instability, particularly at small sample sizes and RF struggled consistently, likely due to slower convergence in limited samples.

In contrast, TDA models remained stable and accurate across all marginal distributions. Their parametric flexibility and faster convergence resulted in low-variance and reasonable AUCs even in small samples. The sTDA and lsTDA variants exhibited slightly higher bias due to additional parameterization.
	
\subsubsection{Scenario C: Disease-specific dependence}

Methods assuming a shared correlation structure suffered when disease-specific dependence was present, especially methods like LR, similar to prior findings~\citep{yan2018combining}. Disease-specific TDA models (TDA$_d$, lTDA$_d$, lsTDA$_d$) performed well, whilst models without disease-specific dependence had more bias. LR and GAM, often perceived as assumption-agnostic, underperformed. This highlights that they implicitly assume equal dependence structures across disease populations.

\subsubsection{Scenario D: Tail dependence}

None of the methods fully captured tail dependence. 
Tree-based methods (RF, XGBoost) struggled the most, exhibiting high variance and bias. Among the TDA variants, location-only and location-scale versions (TDA, lsTDA) were most robust, showing moderate bias and consistent performance. This suggests that TDA’s parametric structure tolerates modest misspecification better than black-box alternatives, especially at small sample sizes.

\subsubsection{Scenario E: Logistic model}

In the correctly specified linear logistic model, LR and LDA unsurprisingly had the lowest bias. TDA followed closely. The sTDA and lsTDA variants exhibited slightly higher bias due to additional parameterization but remained more stable than flexible competitors like GAM, RF, and XGBoost. In the interaction setting, all methods experienced performance drops, with LR and LDA affected most due to model misspecification. TDA variants showed reasonable robustness, particularly lsTDA and lsTDA$_d$, which maintained lower variance and modest bias. This suggests that even when misspecified, TDA can model complex interactions and dependence structures.
 
\section{Optimal diagnostic test for hepatocellular carcinoma}
\label{sec:application}

\subsection{Serum biomarker data and multivariate model}

We analyzed published data from a case-control study involving $N
= 401$ subjects, consisting of $N_1 = 208$ subjects with hepatocellular carcinoma (HCC)
and nondiseased group of $N_0 = 193$ subjects diagnosed with liver cirrhosis, all of whom
exhibited viral or non-viral etiology~\citep{jang2016diagnostic,jang2016diagnostic_data}. The diagnosis of HCC and liver cirrhosis
was established through histological examinations. Imaging studies were
conducted on patients with liver cirrhosis to exclude hepatocellular carcinoma.

Plasma samples from these subjects were analyzed for multiple biomarkers,
including alpha-fetoprotein (AFP), protein induced by vitamin K absence or
antagonist-II (PIVKA-II), osteopontin (OPN), and Dickkopf-1 (DKK-1). While AFP
stands as the most established biomarker for HCC diagnosis, its standalone
diagnostic performance can fall short \citep{tateishi2008diagnostic}. The goal
of our analysis was to create an optimal diagnostic test using a combination of
the measured biomarkers for the detection of HCC and thereby capturing different
aspects of HCC heterogeneity.

Given the right-skewed distributions of all markers, we initially performed a
logarithmic transformation of their measurement values. Unlike our
TDA approach or tree-based (random forest
and boosting) methods, other candidate methods are not invariant to such monotone transformations
and will benefit from more symmetric marginal biomarker distributions.
We assessed the various complexities of our methods
and compared them with competitor methods, detailed in~Section
\ref{sec:empeval}. We used a repeated holdout validation procedure with a 50-50 data splits over 1{,}000 replications to estimate the out-of-sample (OOS) AUC and compare competing methods. The resulting empirical out-of-sample AUCs are shown in Appendix~\ref{sec:mod_sel} Figure~\ref{fig:oosauc}. The multivariate transformation model with
location-scale marginal models and a global covariance matrix (lsTDA) yielded
the highest median empirical out-of-sample AUC, leading us to select it for the subsequent analysis.

Table~\ref{tab:coefs} displays the coefficient estimates from the lsTDA
multivariate transformation model for the HCC biomarkers (DKK-1, OPN, PIVKA-II,
AFP) along with their corresponding 95\% confidence intervals (computed by a
parametric bootstrap). The estimated marginal transformation functions for each
of the biomarkers are provided in Section~\ref{sec:add} Figure~\ref{fig:trafo}
of the supplementary materials. Positive location terms signify that individuals
with HCC exhibit higher biomarker values compared to those without HCC, with the
magnitude indicating the strength of the location shift. Positive scale terms
suggest that biomarker values for HCC subjects display greater variability than
those without HCC. This pattern holds true for all biomarkers, indicating
consistently elevated biomarker measurements and increased variability in
subjects with HCC.
\begin{table}[t!]
	\centering
	\begin{tabular}{lr}
		\toprule 
		{Variable} & {Coefficient ($95$\% CI)} \\
		\midrule 
		{Location} $\delta_j$ & \\
		\quad DKK-1 & 0.721 (0.471, 0.935) \\ 
		\quad OPN & 0.780 (0.443, 1.059) \\ 
		\quad PIVKA-II & 1.257 (0.982, 1.527) \\ 
		\quad AFP & 1.572 (1.262, 1.859) \\ 
		{Scale} $\gamma_j$ & \\
		\quad DKK-1 & 0.499 (0.232, 0.762) \\ 
		\quad OPN & 1.232 (0.987, 1.553) \\ 
		\quad PIVKA-II & 0.694 (0.444, 0.974) \\ 
		\quad AFP & 0.753 (0.495, 1.060) \\ 
		{Correlation} $\mSigma$ & \\
		\quad OPN - DKK-1 & 0.104 (0.011, 0.187) \\ 
		\quad PIVKA-II - DKK-1 & 0.302 (0.210, 0.375) \\ 
		\quad PIVKA-II - OPN & 0.315 (0.227, 0.404) \\ 
		\quad AFP - DKK-1 & 0.232 (0.141, 0.310) \\ 
		\quad AFP - OPN & 0.348 (0.256, 0.421) \\ 
		\quad AFP - PIVKA-II & 0.833 (0.789, 0.861) \\ 
		\bottomrule 
	\end{tabular}
	\caption{Estimated coefficients of the multivariate transformation model with 
		location-scale marginals and global correlation matrix (lsTDA), along with their corresponding 95\% confidence 
		intervals (CI), for the biomarkers employed in 
		hepatocellular carcinoma diagnosis.} 
	\label{tab:coefs}
\end{table}

\subsection{Model-based ROC curves and AUC}

\begin{figure}
	\centering
	\includegraphics[height=13cm, keepaspectratio]{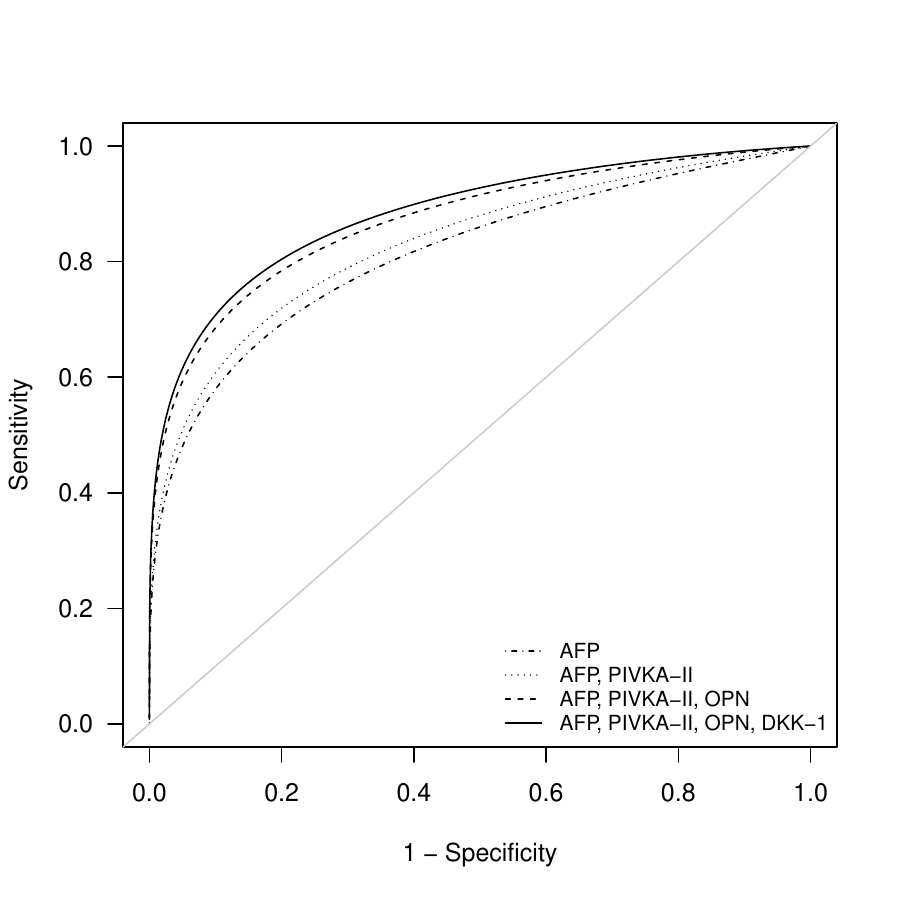} 
	\caption{Estimated model-based ROC curves for the cumulative diagnostic benefit of adding biomarkers to AFP for hepatocellular carcinoma diagnosis.}
	\label{fig:cum_roc}
\end{figure}

Figure~\ref{fig:cum_roc} displays the estimated optimal model-based ROC curves resulting from likelihood ratio combinations of biomarker subsets. AFP was placed first, as it is the most commonly used marker in diagnostic studies of HCC. Additional biomarkers were added sequentially in order of increasing marginal AUC. While this ordering was arbitrary, Table~\ref{tab:aucs} presents results for all possible permutations to provide a comprehensive evaluation. Each subset yields a distinct optimal ROC curve, and as expected, diagnostic accuracy generally improves with the inclusion of additional biomarkers. Note that all combinations can be evaluated from the same fitted model, enabling efficient assessment of multiple diagnostic strategies without refitting.

Table~\ref{tab:aucs} reports both model-based AUCs with 95\% confidence intervals and the corresponding mean out-of-sample (OOS) AUCs. AFP alone had a model-based AUC of 0.814 (95\% CI: 0.769 to 0.850) and an OOS AUC of 0.775. The full four-biomarker model achieved the highest performance, with a model-based AUC of 0.883 (95\% CI: 0.854 to 0.912) and a mean OOS AUC of 0.831, highlighting the value of combining markers. Among three-biomarker combinations, those including AFP, OPN, and either DKK-1 or PIVKA-II achieved nearly comparable model-based AUCs of 0.872 and relatively high OOS AUCs of 0.826. This demonstrates that meaningful gains can be made even without using all four biomarkers.
\begin{table}
	\centering
	\begin{tabular}{lcc}
	\toprule	
	Combination & AUC (95\% CI) & Mean OOS AUC\\ 
	\midrule
	AFP & 0.814 (0.773, 0.850) & 0.775 (0.728, 0.823) \\ 
	PIVKA-II & 0.767 (0.721, 0.809) & 0.721 (0.671, 0.775) \\ 
	OPN & 0.716 (0.678, 0.753) & 0.690 (0.636, 0.742) \\ 
	DKK-1 & 0.674 (0.628, 0.719) & 0.656 (0.606, 0.707) \\ 
	AFP \& PIVKA-II & 0.832 (0.793, 0.866) & 0.782 (0.731, 0.827) \\ 
	AFP \& OPN & 0.858 (0.825, 0.887) & 0.815 (0.766, 0.858) \\ 
	AFP \& DKK-1 & 0.833 (0.797, 0.870) & 0.801 (0.756, 0.849) \\ 
	AFP \& PIVKA-II \& OPN & 0.871 (0.840, 0.898) & 0.820 (0.777, 0.864) \\ 
	AFP \& PIVKA-II \& DKK-1 & 0.849 (0.815, 0.883) & 0.805 (0.760, 0.849) \\ 
	AFP \& OPN \& PIVKA-II & 0.871 (0.840, 0.898) & 0.821 (0.776, 0.865) \\ 
	AFP \& OPN \& DKK-1 & 0.872 (0.841, 0.900) & 0.826 (0.784, 0.867) \\ 
	AFP \& DKK-1 \& PIVKA-II & 0.849 (0.815, 0.883) & 0.806 (0.761, 0.846) \\ 
	AFP \& DKK-1 \& OPN & 0.872 (0.841, 0.900) & 0.826 (0.783, 0.867) \\ 
	AFP \& PIVKA-II \& OPN \& DKK-1 & 0.883 (0.855, 0.910) & 0.832 (0.789, 0.873) \\
	\bottomrule
	\end{tabular}
	\caption{Estimated optimal model-based and out-of-sample AUCs for the likelihood ratio combination of biomarkers, along with their corresponding 95\% confidence intervals (CI).}
	\label{tab:aucs}
\end{table}

\begin{figure}[t!]
	\centering
	\includegraphics[width=\linewidth]{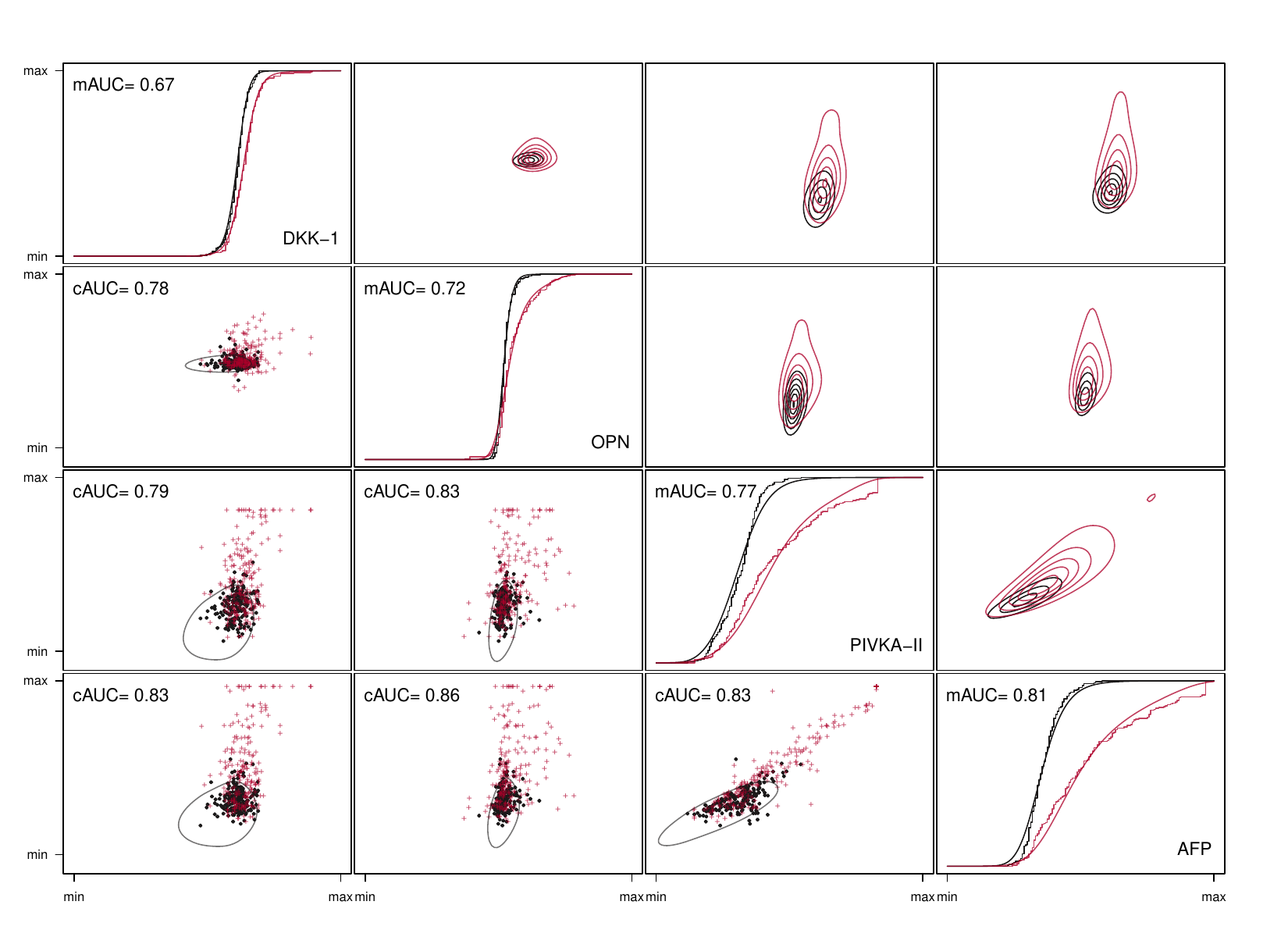} 
	\caption{Visualization of a multivariate transformation model with location-scale marginals and global correlation matrix (lsTDA). 
        Diagonal: Empirical (step) and modeled (smooth) marginal
	distribution functions for each biomarker. The marginal model-based AUC (mAUC) for
	diagnosing HCC using each individual biomarker is provided in the top left
	of each panel. Lower off-diagonal: Bivariate scatterplot of biomarker
	combinations with subjects without HCC (denoted by points $\Large \cdot$) and
	with HCC (denoted by $\color{red} {+}$). The gray line
	represents the modeled likelihood ratio function decision boundary. The
	cumulative model-based AUC (cAUC) for each optimal bivariate biomarker combination is
	provided in the top left of each panel. Upper off-diagonal: Estimated
	bivariate density function for each biomarker combination. In all plots,
	black denotes no HCC, and red signifies HCC.}
	\label{fig:bivariate}
\end{figure}

We also explored all pairwise combinations, as depicted in
Figure~\ref{fig:bivariate}. On the diagonal of this figure, the marginal CDFs estimated using
polynomials in Bernstein form (with $M = 6$) approximate the marginal ECDFs well. However, the marginal models for the HCC subjects do not fit
particularly well for PIVKA-II and AFP due to some extreme measurements. These
observations may be due to upper detection limits for the biomarkers, which
would need to be appropriately addressed in the model by right-censoring, but were beyond
the scope of this analysis.

The lower off-diagonal plots feature two-dimensional scatterplots of the
biomarker data. Recall that the likelihood ratio combination of biomarkers
classifies a subject as an HCC case if their composite score exceeds some cutoff
$c$ (here $\log(L(\yvec)) > 0$ was used), signifying stronger evidence for an
HCC diagnosis. The gray line marks the decision boundary of the modeled
likelihood ratio function in the two-dimensional marker space under this rule.
The most effective bivariate combination involves OPN and AFP, nearly reaching
the same cAUC as using all four markers.

\subsection{Covariate dependent analysis}
\begin{figure}[t!]
	\centering
	\includegraphics[width=\linewidth]{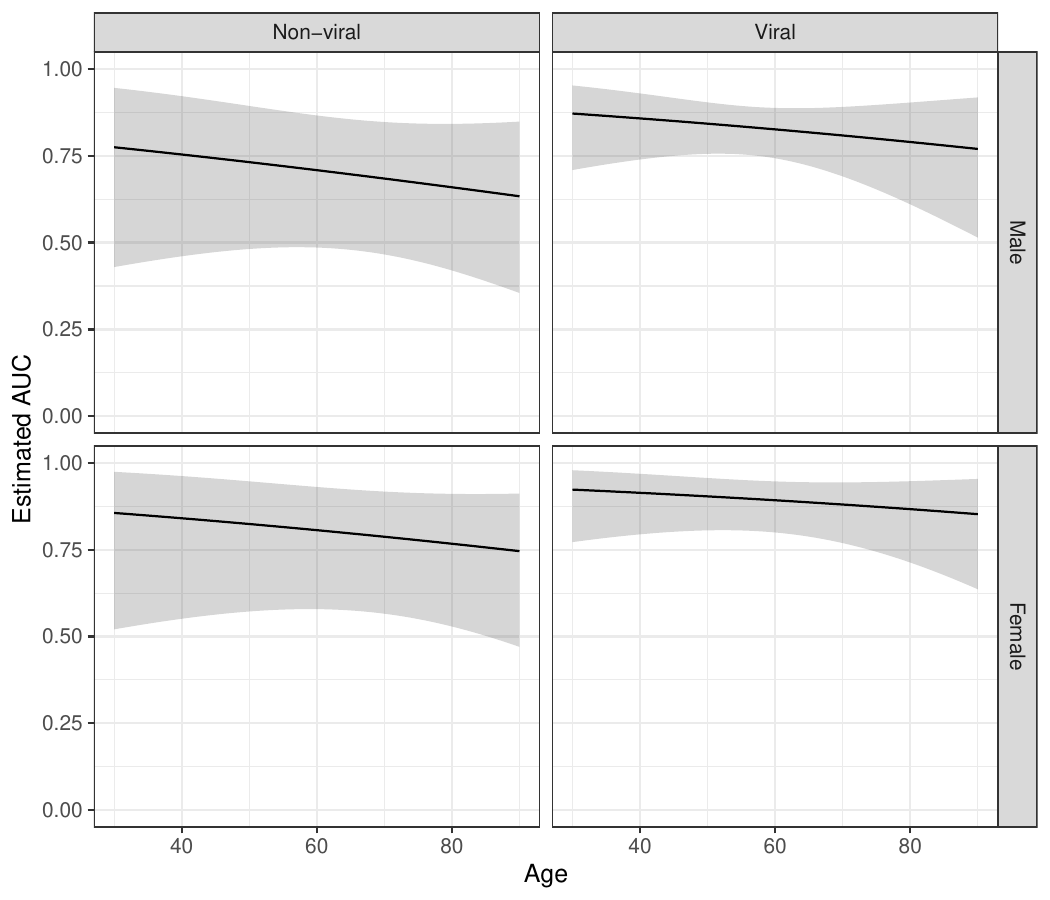} 
	\caption{Estimated covariate-dependent AUCs using the composite OOS likelihood
	ratio score of a multivariate transformation model with 
        location-scale marginals and global correlation matrix (lsTDA), 
        segmented by age and etiological groups, distinguishing between
	viral causes (HBV, HCV) and other etiologies such as alcohol-related or
	cryptogenic factors.}
	\label{fig:auc_cov}
\end{figure}

In their initial analysis of the data, it was observed that
covariates such as age, gender and HCC etiology influenced the individual
diagnostic performance of biomarkers~\citep{jang2016diagnostic}. To evaluate how these factors affect the
diagnostic accuracy of the composite score, we employed a covariate-specific AUC
model defined by
\begin{align*}
	\AUC(\rx) = \frac{\exp(\delta(\rx)) \left( \exp(\delta(\rx) - 1 - \delta(\rx))\right)}{\left(\exp(\delta(\rx)) - 1 \right)^2},
\end{align*}
where $\rx$ represents the covariates age, gender, and etiology, and
$\delta(\rx)$ denotes the covariate effect on the ROC curve. Briefly,
$\text{AUC}(\rx)$ arises from a univariate lTDA model with $F =
\text{logit}^{-1}$ featuring a covariate-dependent location
term $\delta(\rx)$, further details are available in \citep{sewak2023estimating}.
To ensure unbiased AUCs, we initially
computed an average out-of-sample log-likelihood ratio score and then examined
the covariates' dependence on this score. The results are depicted in
Figure~\ref{fig:auc_cov}. For comparison, we used a random forest to generate a
similar score based on the conditional class probability, yielding results
consistent with our method (Figure~\ref{fig:auc_covRF}). The composite score shows higher diagnostic accuracy
for younger ages with gender not having a substantial impact. Furthermore, our
composite score improves the accuracy of HCC detection for viral etiologies,
despite the documented lower accuracy of AFP in identifying viral-related
HCC~\citep{johnson2001role, gopal2014factors}.  This improvement is likely a
result of the complementary information from other markers within the likelihood
ratio combination.

\section{Discussion}
\label{sec:discussion}

Accurate diagnostic tests are essential for routine surveillance and timely
identification of diseases. In this article, we proposed a multivariate modeling framework called Transformation Discriminant Analysis (TDA) to combine multiple biomarkers using the likelihood ratio function. TDA offers flexibility and allows modeling of key clinical complexities such as skewed marginals, disease-specific dependenceand missing biomarkers. Its parametric form enables likelihood-based inference for model parameters and diagnostic accuracy metrics.

Across a range of simulation scenarios, TDA consistently provided accurate diagnostic scores. In settings with skewed marginals or disease-specific dependence structures, commonly used methods such as logistic regression and discriminant analysis showed notable degradation in performance. Machine learning approaches like random forests and GAMs also struggled, especially in small sample sizes, exhibiting high variability and slower convergence. In contrast, TDA models demonstrated lower variance and better adaptability to distributional shifts. In disease-specific dependence scenarios, the advantage of correctly specifying correlation structures was evident. Even under model misspecification, such as tail-dependent copula structures or when the true model followed a logistic regression, TDA variants remained competitive.

When modeling the dependence structure of multivariate data using a parametric copula family, such as the Gaussian copula used in our approach, a common challenge lies in the potential misspecification of the copula family. While we explored the severity of this issue through simulations with tail-dependent copulas, our evaluation remains confined to parametric settings. In this context, if any of the conditional regressions lacks monotonicity, the reliability of a copula model for describing the joint distribution diminishes~\citep{dette2014some} . Fortunately, this assumption can be empirically verified by plotting estimated conditional transformations, as done in Figures~\ref{fig:tramme0} and~\ref{fig:tramme1}, which revealed largely monotonic behavior.

We further assessed model fit using the multivariate probability integral transform~\citep{rosenblatt1952remarks}. The results (Appendix~\ref{sec:mod_ass}) indicated good overall fit, though minor issues were detected in modeling AFP values near the upper limit of detection. These could be addressed by treating such observations as right-censored or adapting more flexible marginal models.

When dealing with a high dimensional set of biomarkers ($J \gg N$), the
covariance matrices in our procedure do not have full rank, leading to
inaccurate inference from the estimation process. A potential extension to TDA involves fitting flexible univariate transformation models, mapping observations to an approximately normal scale and applying penalized covariance estimation techniques—similar in spirit to the nonparanormal model~\citep{Lafferty_Liu_Wasserman_2012}.  Because the penalty appears symmetrically in the likelihood of the diseased and nondiseased populations, it cancels out in the likelihood ratio. However, this approach warrants further investigation to assess its validity and practical performance in high-dimensional applications.

While TDA provides a powerful framework for biomarker combinations, it is not a replacement for all existing methods. Discriminative approaches like logistic regression or methods which optimize empirical performance metrics may outperform in scenarios where the likelihood ratio is difficult to estimate accurately or where our model-based assumptions do not hold. Instead, TDA complements these methods by offering a generative perspective with interpretable components, theoretical optimality guarantees and built-in mechanisms for model assessment. Unlike black-box machine learning models, TDA facilitates inspection of estimated distributions and transformations. This can help practitioners understand the contribution of individual biomarkers and evaluate modeling assumptions. This transparency can guide further model refinement and inform clinical decisions.

A reference implementation of transformation discriminant analysis is available
in the \pkg{tram} add-on package \citep{pkg:tram} to the \proglang{R} system for
statistical computing. The empirical results presented in Sections~\ref{sec:empeval} and
\ref{sec:application} can be reproduced by
\begin{knitrout}
\definecolor{shadecolor}{rgb}{0.969, 0.969, 0.969}\color{fgcolor}\begin{kframe}
\begin{alltt}
\hlkwd{install.packages}\hldef{(}\hlsng{"tram"}\hldef{)}
\hlkwd{library}\hldef{(}\hlsng{"tram"}\hldef{)}
\hlkwd{demo}\hldef{(}\hlsng{"hcc"}\hldef{,} \hlkwc{package} \hldef{=} \hlsng{"tram"}\hldef{)}
\end{alltt}
\end{kframe}
\end{knitrout}

\bibliography{refs,packages}

\clearpage

\makeatletter
\renewcommand\thetable{S\@arabic\c@table}
\renewcommand \thefigure{S\@arabic\c@figure}
\renewcommand{\theHtable}{S\arabic{table}}
\renewcommand{\theHfigure}{S\arabic{figure}}
\makeatother
\setcounter{table}{0}
\setcounter{figure}{0}

\begin{appendix}

\section{Computational details}

We used \pkg{mvtnorm} \citep[][version
1.3.3]{pkg:mvtnorm} to sample from multivariate
normal distributions, \pkg{pracma} \citep[][version
2.4.4]{pkg:pracma} for numerical integration of ROC
curves, and \pkg{qrng} \citep[][version
0.0.10]{pkg:qrng} for random number generation.
Furthermore, we used \pkg{tramME} \citep[][version
1.0.7]{pkg:tramME} for estimating conditional
distributions, \pkg{mgcv} \citep[][version 1.9.3]{pkg:mgcv}
for evaluating the generalized chi-square distribution, and \pkg{copula} \citep[][version
1.1.6]{pkg:copula} for simulating from copulas with
tail-dependence.

In evaluating competitor methods, we used \pkg{randomForest} \citep[][version
4.7.1.2]{pkg:randomForest} with 5000 trees, each
with a minimum of 10 observations in terminal nodes and fitted gradient boosting
models with \pkg{xgboost} \citep[][version
1.7.11.1]{pkg:xgboost} where the iterations were
selected by 5-fold cross-validation. For linear and quadratic discriminant
analysis, we used \pkg{MASS} \citep[][version
7.3.65]{pkg:MASS} and mixture and flexible discriminant
analysis were performed using \pkg{mda} \citep[][version
0.5.5]{pkg:mda}.

All computations were performed using \textsf{R} version
4.5.1
\citep{R}.
 
\section{Proofs}
\label{sec:proof}
We restate and present the proofs of our propositions and their corollaries in this section. Proposition~\ref{prop:TDA} gives the log-likelihood ratio function which has fully flexible marginal distributions whilst Proposition~\ref{prop:lsTDA} introduces a location-scale simplification of the marginal distributions which leads to distributional results for the scalar composite score, $L_d = L(\mY_d)$ for $d=\{0, 1\}$  in Corollary~\ref{cor:gchisq}. Equivalence of a special case of our model with the results of \cite{su1993linear} is shown in Corollary~\ref{cor:lTDA}.

\TDA*
\begin{proof}
The modeled density function $f_d$ is the multivariate normal density function with a zero mean vector and a correlation matrix $\mSigma_d$, evaluated at the transformed variables $\hvec_d(\yvec)$, and incorporates the product of the derivatives of the transformation functions. It's logarithm is given by
\begin{align*}
	\log(f_d(\yvec)) = -\frac{1}{2} \left( \hvec_d(\yvec)^\top \mSigma_d^{-1} \hvec_d(\yvec) + J \log(2 \pi) + \log(\lvert \mSigma_d \rvert) \right) + \sum_{j=1}^J \log \left( h'_{dj}(y_j) \right).
\end{align*}
The asserted representation follows from substituting into the definition of the log-likelihood ratio 
\begin{align*}
	\log(L(\yvec)) = \log(f_1(\yvec)) - \log(f_0(\yvec)).
\end{align*}
\end{proof}

\lsTDA*
\begin{proof}
Using Proposition~\ref{prop:TDA} with $\hvec_0(\yvec)=\hvec(\yvec)$ and $\hvec_1(\yvec)= \mGamma \left(\hvec(\yvec) - \deltavec \right) $ we have
\begin{align*}
	\log(L(\yvec)) 
	&= -\frac{1}{2} \left( \log \left( \frac{\lvert \mSigma_1 \rvert}{\lvert \mSigma_0 \rvert} \right) 
	+ \hvec(\yvec)^\top \mA \hvec(\yvec) 
	- 2 \deltavec^\top (\mA + \mSigma_0^{-1}) \hvec(\yvec) \right. \\
	&\quad \left. + \deltavec^\top (\mA + \mSigma_0^{-1}) \deltavec \right) 
	- \sum_{j=1}^J \gamma_j \\
	&\quad \text{(The result follows from completing the square)} \\
	&= -\frac{1}{2} \left( \log \left( \frac{\lvert \mSigma_1 \rvert}{\lvert \mSigma_0 \rvert} \right) 
	+ \left( \hvec(\yvec) - \left(\mI + (\mSigma_0 \mA)^{-1} \right) \deltavec \right)^\top 
	\mA \left( \hvec(\yvec) - \left(\mI + (\mSigma_0 \mA)^{-1} \right)\deltavec \right) \right. \\
	&\quad \left. - \deltavec^\top \left(\mA + \mSigma_0^{-1} \right) \mA^{-1} 
	\left(\mA + \mSigma_0^{-1} \right) \deltavec 
	+ \deltavec^\top (\mA + \mSigma_0^{-1}) \deltavec \right) 
	- \sum_{j=1}^J \gamma_j \\
	&= -\frac{1}{2} \left( \hvec(\yvec) - \betavec \right)^\top \mA \left( \hvec(\yvec) - \betavec \right) + c,
\end{align*}
with constant $ c = -\frac{1}{2} \left( \log \left( \frac{\lvert \mSigma_1 \rvert}{\lvert \mSigma_0 \rvert} \right)
- \deltavec^\top \mSigma_0^{-1} (\mI + (\mSigma_0 \mA)^{-1}) \deltavec \right) - \sum_{j=1}^J \gamma_j.$.
\end{proof}

\begin{definition}
Let $X_1,\dots,X_m$ be independent normally distributed variables with $X_i \sim N(\mu_i, 1)$. The distribution of $\sum_{i = 1}^{m} w_i X_i^2$ is called the generalized chi-square distribution with parameter vectors specifying the weights $\wvec = (w_1,\dots,w_m)^\top$, degrees of freedom $\kvec=\onevec \in \RR^m$ and non-centrality terms $\nuvec = (\mu_1^2,\dots,\mu_m^2)^\top$. 
\end{definition}

\begin{lemma}[Quadratic form of the multivariate normal distribution]
Let $L = \mX^\top \mA \mX$ with $\mX \sim N_J(\muvec, \mSigma)$, $\rank(\mA) = J$ and $\mSigma$ symmetric and positive semi-definite. Let the spectral decomposition of $\mSigma^{\frac{1}{2}} \mA \mSigma^{\frac{1}{2}}$ be given by $\mP \mW \mP^\top$. Then $L$ has a generalized chi-square distribution with weights which are the eigenvalues $\wvec = \diag(\mW)$, degrees of freedom $\onevec \in \RR^J$ and non-centrality parameters $\diag(\mP^\top \mSigma^{\frac{1}{2}}\muvec)^2$.
\label{lem:quad}
\end{lemma}
\begin{proof}
Using the spectral decomposition we can rewrite $L$ as
\begin{align*}
	L & =  \mX^\top \mA \mX \\
	& = \mX^\top \mSigma^{-\frac{1}{2}} \mSigma^{\frac{1}{2}} \mA \mSigma^{\frac{1}{2}} \mSigma^{-\frac{1}{2}} \mX \\
	& =  \mX^\top \mSigma^{-\frac{1}{2}} \mP \mW \mP^\top \mSigma^{-\frac{1}{2}} \mX \\
	&= \mQ^\top \mW \mQ \\
	& = \sum_{i=1}^J w_i Q_i^2,
\end{align*}
where $\mQ = \mP^\top \mSigma^{-\frac{1}{2}} \mX \sim N_J( \mP^\top \mSigma^{-\frac{1}{2}} \muvec, \mI)$. Thus, by definition, $L$ has a generalized chi-square distribution with the given parameters.
\end{proof}

\gchisq*
\begin{proof}
From Proposition~\ref{prop:lsTDA} we have that 
\begin{align*}
	L_d & = \log(L(\mY_d)) \\
	& = -\frac{1}{2} \left(\hvec(\mY_d) - \betavec \right)^\top \mA \left(\hvec(\mY_d) - \betavec \right) + c,
\end{align*}
and the distributions of the transformed random vectors in the two classes are
\begin{align*}
	\hvec(\mY_0) \sim N_J(\nullvec, \mSigma_0) \quad \text{and} \quad \hvec(\mY_1) \sim N_J(\deltavec, \mGamma^{-1} \mSigma_1 \mGamma^{-1}).
\end{align*}
The result follows from an application of Lemma~\ref{lem:quad} with $\mX = \hvec(\mY_d) - \betavec$.
\end{proof}

\section{Cost and resource optimization}
\label{sec:cost}
Interest may also lie in finding a balance between cost-effectiveness and
diagnostic accuracy. We can use the property of the multivariate model detailed
in Section~\ref{sec:subset} to formulate this task as an integer optimization
problem. Define the decision variables $\svec = (s_1,\dots,s_J)^\top$ where $s_j
= 0$ or $1$, indicating if a biomarker is rejected or accepted in the overall
biomarker combination. Assume $K$ resources (\eg machines, cost or time) where $a_{kj}$ is the
amount of resource $k$ used on biomarker~$j$ and $b_k$ is the budget for the
$k$th resource. To find the optimal assignment $\svec^*$, we solve the
optimization problem
\begin{align*}
	\textrm{maximize}_{\svec} \quad & \AUC(\svec) = \Phi \left( \sqrt{ \frac{(\deltahatvec \odot \svec)^\top (\diag(\svec) \widehat{\mSigma}^{-1} \diag(\svec)) (\deltahatvec \odot \svec)}{2}} \right) \\
	\textrm{subject to} \quad & \sum_{j=1}^J a_{kj} s_j \leq b_k \quad (k = 1,\dots,K), \\
	\quad & s_j = 0 \text{ or } 1 \quad (j = 1,\dots,J).
\end{align*}
Here, $\odot$ denotes the Hadamard product (element-wise product), and
$\deltahatvec$ and $\widehat{\mSigma}$ are the estimated model coefficients from
a location-only model as presented in the model-based AUC function. The objective is to
maximize the diagnostic accuracy of the biomarker combination without exceeding the
limited availability of any resource $b_k$. The optimization problem can be
solved without needing to re-fit joint distributions for biomarker subsets. Note
that for more complex models outlined previously, the objective function can be
similarly formulated. While these AUC functions might not be readily available
in closed form, they can be computed numerically based on the model parameters.

\section{Simulation results}

\begin{longtable}{lccc}
\toprule
& \multicolumn{3}{c}{{Sample sizes}}\\
\cmidrule(lr){2-4}
& 50 & 100 & 200 \\ 
\addlinespace[0.7em]
\multicolumn{4}{l}{\emph{Scenario A - Multivariate normal}} \\
\midrule
TDA & 0.025 (0.015) & 0.012 (0.007) & 0.006 (0.003) \\ 
TDA\textsubscript{d} & 0.026 (0.015) & 0.012 (0.007) & 0.006 (0.003) \\ 
lsTDA & 0.032 (0.017) & 0.015 (0.008) & 0.007 (0.004) \\ 
lsTDA\textsubscript{d} & 0.033 (0.017) & 0.015 (0.008) & 0.007 (0.004) \\ 
sTDA & 0.051 (0.020) & 0.029 (0.011) & 0.016 (0.006) \\ 
sTDA\textsubscript{d} & 0.052 (0.020) & 0.030 (0.011) & 0.016 (0.006) \\ 
LDA & 0.015 (0.012) & 0.008 (0.006) & 0.004 (0.003) \\ 
QDA & 0.015 (0.012) & 0.008 (0.006) & 0.004 (0.003) \\ 
MDA & 0.041 (0.026) & 0.019 (0.012) & 0.008 (0.005) \\ 
FDA & 0.015 (0.012) & 0.008 (0.006) & 0.004 (0.003) \\ 
LR & 0.015 (0.012) & 0.008 (0.006) & 0.004 (0.003) \\ 
GAM & 0.070 (0.056) & 0.021 (0.022) & 0.009 (0.010) \\ 
RF & 0.073 (0.015) & 0.061 (0.011) & 0.051 (0.008) \\ 
XGBoost & 0.138 (0.029) & 0.117 (0.024) & 0.095 (0.018) \\ 
KT & 0.081 (0.032) & 0.075 (0.022) & 0.069 (0.016) \\ 
LIU & 0.194 (0.013) & 0.190 (0.008) & 0.187 (0.004) \\ 
\addlinespace[0.7em]
\multicolumn{4}{l}{\emph{Scenario B - Multivariate lognormal}} \\ \midrule
TDA & 0.028 (0.016) & 0.014 (0.007) & 0.008 (0.004) \\ 
TDA\textsubscript{d} & 0.030 (0.016) & 0.015 (0.007) & 0.008 (0.004) \\ 
lsTDA & 0.037 (0.018) & 0.019 (0.008) & 0.010 (0.004) \\ 
lsTDA\textsubscript{d} & 0.038 (0.018) & 0.019 (0.008) & 0.010 (0.004) \\ 
sTDA & 0.055 (0.020) & 0.034 (0.011) & 0.020 (0.006) \\ 
sTDA\textsubscript{d} & 0.056 (0.020) & 0.034 (0.011) & 0.020 (0.006) \\ 
LDA & 0.062 (0.024) & 0.051 (0.017) & 0.044 (0.013) \\ 
QDA & 0.093 (0.021) & 0.087 (0.015) & 0.083 (0.012) \\ 
MDA & 0.096 (0.032) & 0.078 (0.024) & 0.065 (0.017) \\ 
FDA & 0.062 (0.024) & 0.051 (0.017) & 0.044 (0.013) \\ 
LR & 0.057 (0.022) & 0.046 (0.013) & 0.039 (0.010) \\ 
GAM & 0.093 (0.047) & 0.057 (0.026) & 0.039 (0.016) \\ 
RF & 0.075 (0.016) & 0.064 (0.012) & 0.053 (0.009) \\ 
XGBoost & 0.145 (0.030) & 0.122 (0.025) & 0.098 (0.019) \\ 
KT & 0.097 (0.028) & 0.090 (0.021) & 0.086 (0.014) \\ 
LIU & 0.204 (0.012) & 0.200 (0.007) & 0.198 (0.004) \\ 
\addlinespace[0.7em]
\multicolumn{4}{l}{\emph{Scenario B - Multivariate different marginals}} \\ \midrule
TDA & 0.024 (0.014) & 0.014 (0.007) & 0.009 (0.003) \\ 
TDA\textsubscript{d} & 0.026 (0.014) & 0.014 (0.007) & 0.009 (0.003) \\ 
lsTDA & 0.029 (0.015) & 0.014 (0.007) & 0.007 (0.004) \\ 
lsTDA\textsubscript{d} & 0.030 (0.016) & 0.014 (0.008) & 0.007 (0.004) \\ 
sTDA & 0.055 (0.021) & 0.034 (0.012) & 0.019 (0.007) \\ 
sTDA\textsubscript{d} & 0.056 (0.021) & 0.034 (0.012) & 0.019 (0.007) \\ 
LDA & 0.030 (0.013) & 0.023 (0.008) & 0.019 (0.004) \\ 
QDA & 0.051 (0.013) & 0.045 (0.008) & 0.041 (0.006) \\ 
MDA & 0.073 (0.029) & 0.051 (0.018) & 0.039 (0.011) \\ 
FDA & 0.030 (0.013) & 0.023 (0.008) & 0.019 (0.004) \\ 
LR & 0.030 (0.013) & 0.023 (0.007) & 0.019 (0.004) \\ 
GAM & 0.087 (0.060) & 0.041 (0.031) & 0.022 (0.014) \\ 
RF & 0.060 (0.013) & 0.053 (0.011) & 0.045 (0.008) \\ 
XGBoost & 0.120 (0.032) & 0.101 (0.023) & 0.083 (0.017) \\ 
KT & 0.049 (0.023) & 0.039 (0.016) & 0.036 (0.012) \\ 
LIU & 0.213 (0.012) & 0.210 (0.007) & 0.208 (0.004) \\ 
\addlinespace[0.7em]
\multicolumn{4}{l}{\emph{Scenario C - Multivariate\textsubscript{d} normal}} \\ \midrule
TDA & 0.079 (0.013) & 0.069 (0.006) & 0.064 (0.003) \\ 
TDA\textsubscript{d} & 0.031 (0.015) & 0.014 (0.007) & 0.007 (0.003) \\ 
lsTDA & 0.064 (0.018) & 0.051 (0.011) & 0.044 (0.007) \\ 
lsTDA\textsubscript{d} & 0.033 (0.015) & 0.015 (0.007) & 0.007 (0.003) \\ 
sTDA & 0.084 (0.020) & 0.065 (0.014) & 0.053 (0.008) \\ 
sTDA\textsubscript{d} & 0.049 (0.018) & 0.026 (0.010) & 0.014 (0.005) \\ 
LDA & 0.073 (0.011) & 0.066 (0.006) & 0.063 (0.003) \\ 
QDA & 0.013 (0.009) & 0.006 (0.004) & 0.003 (0.002) \\ 
MDA & 0.084 (0.029) & 0.059 (0.019) & 0.047 (0.014) \\ 
FDA & 0.073 (0.011) & 0.066 (0.006) & 0.063 (0.003) \\ 
LR & 0.073 (0.011) & 0.066 (0.006) & 0.063 (0.003) \\ 
GAM & 0.124 (0.054) & 0.079 (0.022) & 0.066 (0.008) \\ 
RF & 0.116 (0.016) & 0.096 (0.013) & 0.076 (0.009) \\ 
XGBoost & 0.182 (0.032) & 0.152 (0.028) & 0.118 (0.021) \\ 
KT & 0.119 (0.029) & 0.112 (0.021) & 0.107 (0.015) \\ 
LIU & 0.179 (0.012) & 0.176 (0.005) & 0.174 (0.002) \\ 
\addlinespace[0.7em]
\multicolumn{4}{l}{\emph{Scenario C - Multivariate\textsubscript{d} lognormal}} \\ \midrule
TDA & 0.085 (0.014) & 0.073 (0.007) & 0.068 (0.003) \\ 
TDA\textsubscript{d} & 0.040 (0.017) & 0.022 (0.009) & 0.014 (0.004) \\ 
lsTDA & 0.067 (0.019) & 0.051 (0.012) & 0.044 (0.007) \\ 
lsTDA\textsubscript{d} & 0.042 (0.017) & 0.023 (0.009) & 0.014 (0.004) \\ 
sTDA & 0.088 (0.022) & 0.065 (0.014) & 0.053 (0.009) \\ 
sTDA\textsubscript{d} & 0.059 (0.020) & 0.034 (0.012) & 0.022 (0.006) \\ 
LDA & 0.112 (0.026) & 0.100 (0.018) & 0.094 (0.013) \\ 
QDA & 0.104 (0.026) & 0.097 (0.021) & 0.092 (0.017) \\ 
MDA & 0.149 (0.035) & 0.131 (0.026) & 0.117 (0.019) \\ 
FDA & 0.112 (0.026) & 0.100 (0.018) & 0.094 (0.013) \\ 
LR & 0.108 (0.023) & 0.097 (0.015) & 0.090 (0.011) \\ 
GAM & 0.144 (0.046) & 0.113 (0.027) & 0.096 (0.015) \\ 
RF & 0.120 (0.018) & 0.100 (0.016) & 0.079 (0.011) \\ 
XGBoost & 0.186 (0.034) & 0.157 (0.029) & 0.122 (0.022) \\ 
KT & 0.132 (0.026) & 0.122 (0.020) & 0.118 (0.016) \\ 
LIU & 0.196 (0.011) & 0.192 (0.005) & 0.191 (0.002) \\ 
\addlinespace[0.7em]
\multicolumn{4}{l}{\emph{Scenario C - Multivariate\textsubscript{d} different marginals}} \\ \midrule
TDA & 0.085 (0.015) & 0.074 (0.007) & 0.069 (0.003) \\ 
TDA\textsubscript{d} & 0.036 (0.015) & 0.019 (0.008) & 0.012 (0.004) \\ 
lsTDA & 0.061 (0.017) & 0.047 (0.010) & 0.040 (0.006) \\ 
lsTDA\textsubscript{d} & 0.037 (0.015) & 0.018 (0.008) & 0.010 (0.003) \\ 
sTDA & 0.089 (0.020) & 0.066 (0.014) & 0.051 (0.008) \\ 
sTDA\textsubscript{d} & 0.059 (0.018) & 0.034 (0.012) & 0.019 (0.006) \\ 
LDA & 0.089 (0.014) & 0.081 (0.008) & 0.077 (0.005) \\ 
QDA & 0.069 (0.011) & 0.061 (0.007) & 0.056 (0.004) \\ 
MDA & 0.119 (0.029) & 0.098 (0.018) & 0.085 (0.012) \\ 
FDA & 0.089 (0.014) & 0.081 (0.008) & 0.077 (0.005) \\ 
LR & 0.089 (0.014) & 0.080 (0.007) & 0.077 (0.004) \\ 
GAM & 0.151 (0.058) & 0.104 (0.033) & 0.082 (0.015) \\ 
RF & 0.110 (0.014) & 0.097 (0.012) & 0.083 (0.009) \\ 
XGBoost & 0.171 (0.032) & 0.149 (0.025) & 0.122 (0.018) \\ 
KT & 0.105 (0.022) & 0.094 (0.015) & 0.090 (0.010) \\ 
LIU & 0.254 (0.011) & 0.250 (0.007) & 0.247 (0.003) \\ 
\addlinespace[0.7em]
\multicolumn{4}{l}{\emph{Scenario D - Gumbel copula dependence}} \\ \midrule
TDA & 0.033 (0.015) & 0.023 (0.008) & 0.018 (0.004) \\ 
TDA\textsubscript{d} & 0.034 (0.015) & 0.024 (0.008) & 0.020 (0.004) \\ 
lsTDA & 0.036 (0.018) & 0.019 (0.009) & 0.012 (0.004) \\ 
lsTDA\textsubscript{d} & 0.036 (0.018) & 0.019 (0.009) & 0.012 (0.005) \\ 
sTDA & 0.060 (0.023) & 0.037 (0.013) & 0.024 (0.007) \\ 
sTDA\textsubscript{d} & 0.061 (0.023) & 0.038 (0.013) & 0.024 (0.007) \\ 
LDA & 0.031 (0.016) & 0.021 (0.009) & 0.017 (0.005) \\ 
QDA & 0.048 (0.015) & 0.038 (0.008) & 0.034 (0.005) \\ 
MDA & 0.068 (0.031) & 0.043 (0.019) & 0.030 (0.010) \\ 
FDA & 0.031 (0.016) & 0.021 (0.009) & 0.017 (0.005) \\ 
LR & 0.029 (0.014) & 0.019 (0.007) & 0.015 (0.003) \\ 
GAM & 0.090 (0.061) & 0.040 (0.032) & 0.022 (0.014) \\ 
RF & 0.059 (0.015) & 0.049 (0.011) & 0.041 (0.008) \\ 
XGBoost & 0.119 (0.034) & 0.095 (0.023) & 0.076 (0.016) \\ 
KT & 0.053 (0.024) & 0.040 (0.013) & 0.035 (0.007) \\ 
LIU & 0.218 (0.008) & 0.216 (0.007) & 0.213 (0.004) \\ 
\addlinespace[0.7em]
\multicolumn{4}{l}{\emph{Scenario D - Clayton copula dependence}} \\ \midrule
TDA & 0.029 (0.013) & 0.019 (0.006) & 0.015 (0.003) \\ 
TDA\textsubscript{d} & 0.029 (0.014) & 0.018 (0.006) & 0.013 (0.003) \\ 
lsTDA & 0.035 (0.016) & 0.020 (0.007) & 0.014 (0.004) \\ 
lsTDA\textsubscript{d} & 0.036 (0.016) & 0.020 (0.007) & 0.014 (0.004) \\ 
sTDA & 0.060 (0.019) & 0.040 (0.013) & 0.027 (0.007) \\ 
sTDA\textsubscript{d} & 0.060 (0.019) & 0.040 (0.013) & 0.026 (0.007) \\ 
LDA & 0.042 (0.013) & 0.034 (0.007) & 0.030 (0.004) \\ 
QDA & 0.065 (0.013) & 0.058 (0.007) & 0.055 (0.005) \\ 
MDA & 0.082 (0.027) & 0.061 (0.018) & 0.049 (0.011) \\ 
FDA & 0.042 (0.013) & 0.034 (0.007) & 0.030 (0.004) \\ 
LR & 0.041 (0.013) & 0.032 (0.006) & 0.029 (0.003) \\ 
GAM & 0.105 (0.059) & 0.051 (0.031) & 0.033 (0.015) \\ 
RF & 0.051 (0.012) & 0.045 (0.010) & 0.040 (0.008) \\ 
XGBoost & 0.114 (0.033) & 0.095 (0.024) & 0.077 (0.017) \\ 
KT & 0.050 (0.018) & 0.040 (0.009) & 0.036 (0.005) \\ 
LIU & 0.202 (0.012) & 0.199 (0.006) & 0.198 (0.004) \\ 
\addlinespace[0.7em]
\multicolumn{4}{l}{\emph{Scenario E - linear logistic model}} \\ \midrule
TDA & 0.034 (0.028) & 0.016 (0.012) & 0.008 (0.006) \\ 
TDA\textsubscript{d} & 0.075 (0.042) & 0.037 (0.021) & 0.016 (0.009) \\ 
lsTDA & 0.067 (0.040) & 0.032 (0.019) & 0.015 (0.009) \\ 
lsTDA\textsubscript{d} & 0.093 (0.045) & 0.049 (0.024) & 0.023 (0.011) \\ 
sTDA & 0.104 (0.044) & 0.064 (0.025) & 0.036 (0.014) \\ 
sTDA\textsubscript{d} & 0.117 (0.046) & 0.075 (0.027) & 0.043 (0.015) \\ 
LDA & 0.031 (0.027) & 0.015 (0.012) & 0.007 (0.006) \\ 
QDA & 0.082 (0.042) & 0.045 (0.023) & 0.021 (0.010) \\ 
MDA & 0.092 (0.041) & 0.055 (0.025) & 0.028 (0.012) \\ 
FDA & 0.031 (0.027) & 0.015 (0.012) & 0.007 (0.006) \\ 
LR & 0.032 (0.028) & 0.015 (0.012) & 0.007 (0.006) \\ 
GAM & 0.119 (0.069) & 0.048 (0.040) & 0.019 (0.016) \\ 
RF & 0.076 (0.031) & 0.061 (0.020) & 0.048 (0.013) \\ 
XGBoost & 0.123 (0.043) & 0.101 (0.030) & 0.082 (0.020) \\ 
KT & 0.303 (0.004) & 0.303 (0.005) & 0.304 (0.006) \\ 
LIU & 0.306 (0.002) & 0.307 (0.002) & 0.308 (0.002) \\ 
\addlinespace[0.7em]
\multicolumn{4}{l}{\emph{Scenario E - logistic model with interactions}} \\ \midrule
TDA & 0.121 (0.033) & 0.102 (0.016) & 0.092 (0.008) \\ 
TDA\textsubscript{d} & 0.131 (0.039) & 0.094 (0.021) & 0.075 (0.011) \\ 
lsTDA & 0.099 (0.040) & 0.064 (0.020) & 0.045 (0.010) \\ 
lsTDA\textsubscript{d} & 0.102 (0.041) & 0.056 (0.022) & 0.030 (0.011) \\ 
sTDA & 0.140 (0.043) & 0.099 (0.025) & 0.070 (0.014) \\ 
sTDA\textsubscript{d} & 0.134 (0.041) & 0.086 (0.025) & 0.052 (0.013) \\ 
LDA & 0.119 (0.032) & 0.102 (0.016) & 0.092 (0.007) \\ 
QDA & 0.097 (0.039) & 0.055 (0.021) & 0.029 (0.011) \\ 
MDA & 0.117 (0.040) & 0.073 (0.024) & 0.043 (0.013) \\ 
FDA & 0.119 (0.032) & 0.102 (0.016) & 0.092 (0.007) \\ 
LR & 0.119 (0.032) & 0.102 (0.016) & 0.092 (0.007) \\ 
GAM & 0.157 (0.066) & 0.085 (0.038) & 0.054 (0.017) \\ 
RF & 0.110 (0.033) & 0.081 (0.019) & 0.063 (0.011) \\ 
XGBoost & 0.160 (0.046) & 0.126 (0.029) & 0.097 (0.018) \\ 
KT & 0.339 (0.003) & 0.339 (0.003) & 0.340 (0.003) \\ 
LIU & 0.333 (0.003) & 0.333 (0.003) & 0.332 (0.003) \\ 
\bottomrule
\caption{Bias and standard error (in parentheses) of out-of-sample AUC estimates for different methods. Results are stratified by sample size and simulation scenarios. Values represent averages over 1,000 simulation replicates.}
\label{tab:sim}
\end{longtable}

\section{Model selection and assessment}
\label{sec:mod_sel}
\subsection{Model selection}

To determine the required model flexibility and evaluate performance against competing methods for hepatocellular carcinoma diagnosis we employed a repeated holdout validation procedure. This involved comparing various methods suitable for combining multiple biomarkers into an optimal diagnostic test. We computed the out-of-sample AUC for each method for the hepatocellular carcinoma (HCC) data. Details on these methods is given in Section~\ref{sec:empeval} and on the dataset in Section~\ref{sec:application} of the main text.

The repeated holdout validation procedure is an unbiased procedure for model selection and minimizes the bias associated with holdout validation. Briefly, the steps of the procedure are as follows:
\begin{enumerate}
	\item Randomly divide the data into two subsets of equal size: a training and holdout set.
	\item For each model, estimate its parameters from the training set and using this fitted model calculate the composite score (log-likelihood ratio or the positive class probability) on the holdout set.
	\item Repeat steps 1 and 2 to obtain a distribution of out-of-sample AUCs
	(we used $1000$ iterations).
\end{enumerate}

\begin{figure}
	\centering
	\includegraphics[width=\linewidth]{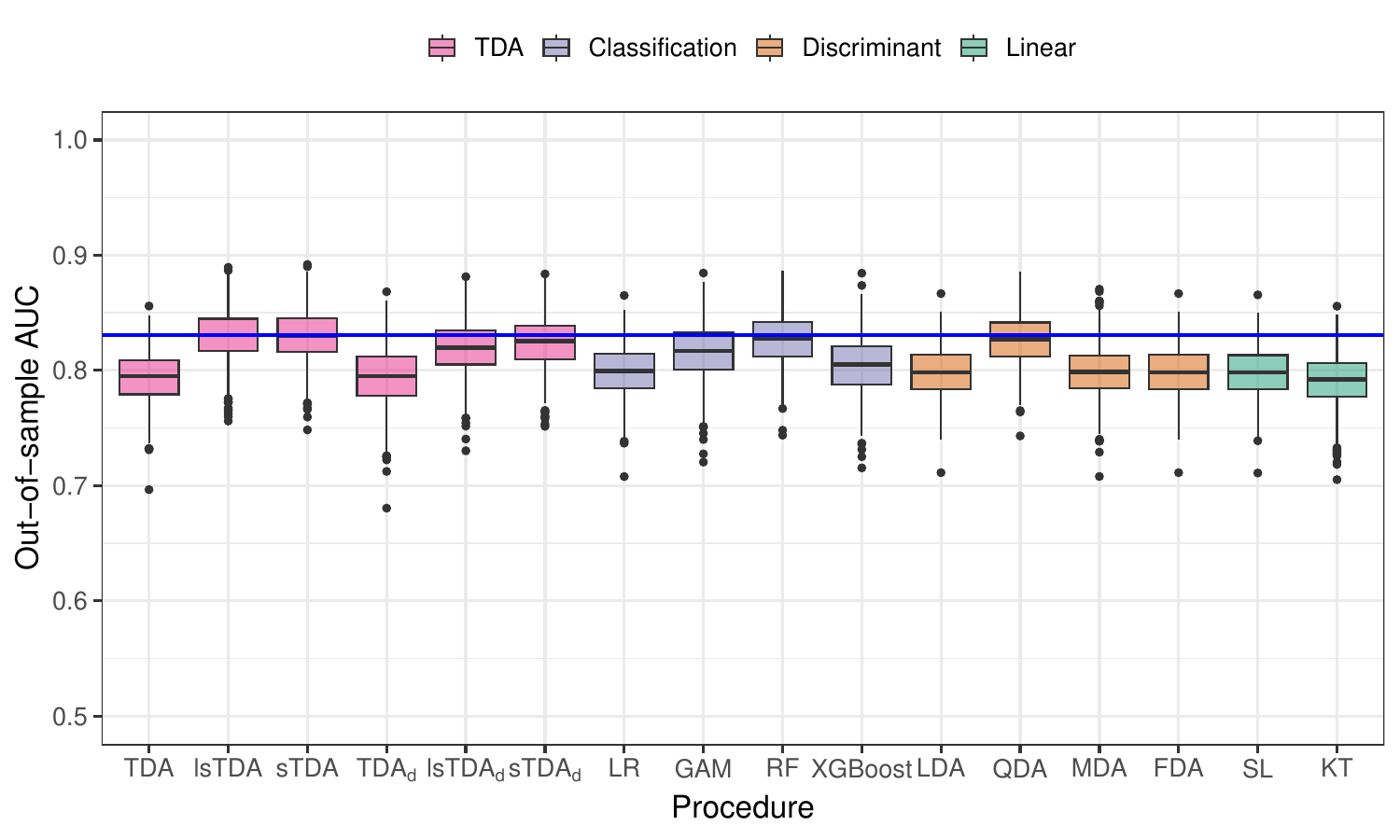} 
	\caption{Distributions of out-of-sample (OOS) area under the curve (AUC) values obtained by combining four biomarkers (AFP, PIVKA-II, OPN, and DKK-1) to generate an optimal diagnostic score. The box-plots are color-coded to categorize methods with those using the multivariate transformation model (TDA), and competitors including binary classification, discriminant analysis and linear combination methods. The blue line indicates the maximum median OOS AUC which was for the lsTDA method.}
	\label{fig:oosauc}
\end{figure}

The results depicted in Figure~\ref{fig:oosauc} reveal consistent performance across all methods, as indicated by
out-of-sample (OOS) AUC quartiles ranging between $0.75$ to $0.85$. As discussed in the main text, the variance in the biomarker distributions among HCC cases was higher than among liver cirrhosis cases (control). Consequently, the subset of proposed methods (lTDA, lTDA$_d$) with only a location parameter had a relatively lower performance compared to methods that incorporated scale. Similarly, since QDA can capture different marginal variances it also performed well whilst other discriminant analysis or linear combination methods lacked this capability.  The random forest classification method demonstrated good performance, aligning with findings from the simulation study. The multivariate transformation model with location-scale marginal models and a global covariance matrix (lsTDA) yielded the highest median OOS AUC, leading us to select it for the subsequent analysis.

\subsection{Model assessment}
\label{sec:mod_ass}
\subsubsection{Rosenblatt transformation}
For a univariate distribution $Y \in \RR$, a model diagnostic technique involves plotting the empirical CDF of the probability integral transform (PIT) values and contrasting it with the CDF of the uniform distribution. Analogously, the dependent random vector $\mY = (Y_1, \dots, Y_J)^\top \in \RR^J$ can be transformed into a uniform random vector $\mU = (U_1,\dots, U_J)^\top$ by the transformation of \cite{rosenblatt1952remarks}. Each of the margins $U_j$ are independent $\text{Uniform}(0,1)$ random variables. This transformation is given by
\begin{align*}
	U_1 & = F_1(Y_1) \\
	U_2 & = F_2(Y_2 \mid Y_1) \\
	\vdots \\
	U_J & = F_J(Y_J \mid Y_1, \dots, Y_{J-1})
\end{align*}
where the conditional CDF are $F_j(y_j \mid \yvec_{j-1}) = \Prob(Y_j \leq y_j \mid \mY_{j-1} = \yvec_{j-1})$.

Our multivariate transformation model assumes that the \emph{transformed} random vector has a multivariate normal distribution. Under this model, by the properties of the 
multivariate normal distribution, the conditional distributions have a univariate normal distribution, with mean and covariance structures depending on the model parameters. Thus, we can
derive estimates of these conditional CDFs by
\begin{align*}
	\hat{F}_j(y_j \mid \yvec_{j-1}) &= \hat{\Prob}(Y_j \leq y_j \mid \mY_{j-1} = \yvec_{j-1}) 
	= \Phi \left(\sum_{k = 1}^j \hat{\tilde{\mLambda}}_{jk} \hhat_{j}(y_{j})\right)
\end{align*}
where $\hhat_j(y_j)$ represents the estimated transformation functions for $j=1,\dots, J$,
evaluated at the $j$th biomarker value and $\hat{\tilde{\mLambda}}_{jk}$ is the element in the $j$th row and $k$th column of
the estimate of $\tilde{\mLambda}$ our model is parameterized with.
Consequently, we employ the Rosenblatt transformation, using the modeled conditional distribution functions to map the original biomarker measurements to uniformity. 

\begin{figure}
	\centering
	\includegraphics[width=\linewidth]{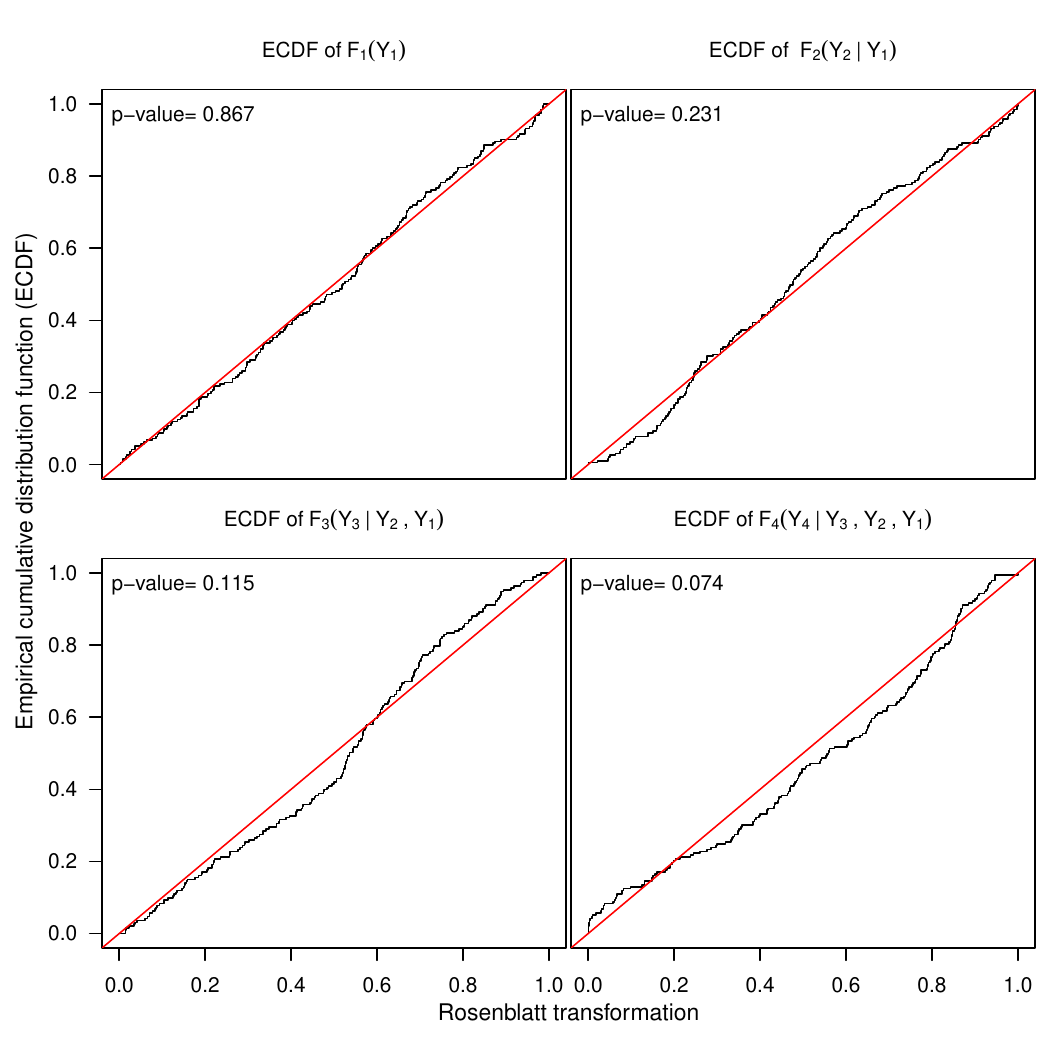} 
	\caption{Model diagnostic plots for the multivariate transformation model with 
                location-scale marginals and global correlation matrix (lsTDA) in subjects without hepatocellular carcinoma, employing the Rosenblatt transformation. The black line represents the empirical cumulative distribution function (ECDF) of the marginal Rosenblatt transformed data, while the red line represents the theoretical CDF of a uniform distribution.}
	\label{fig:mod_gof0}
\end{figure}

\begin{figure}
	\centering
	\includegraphics[width=\linewidth]{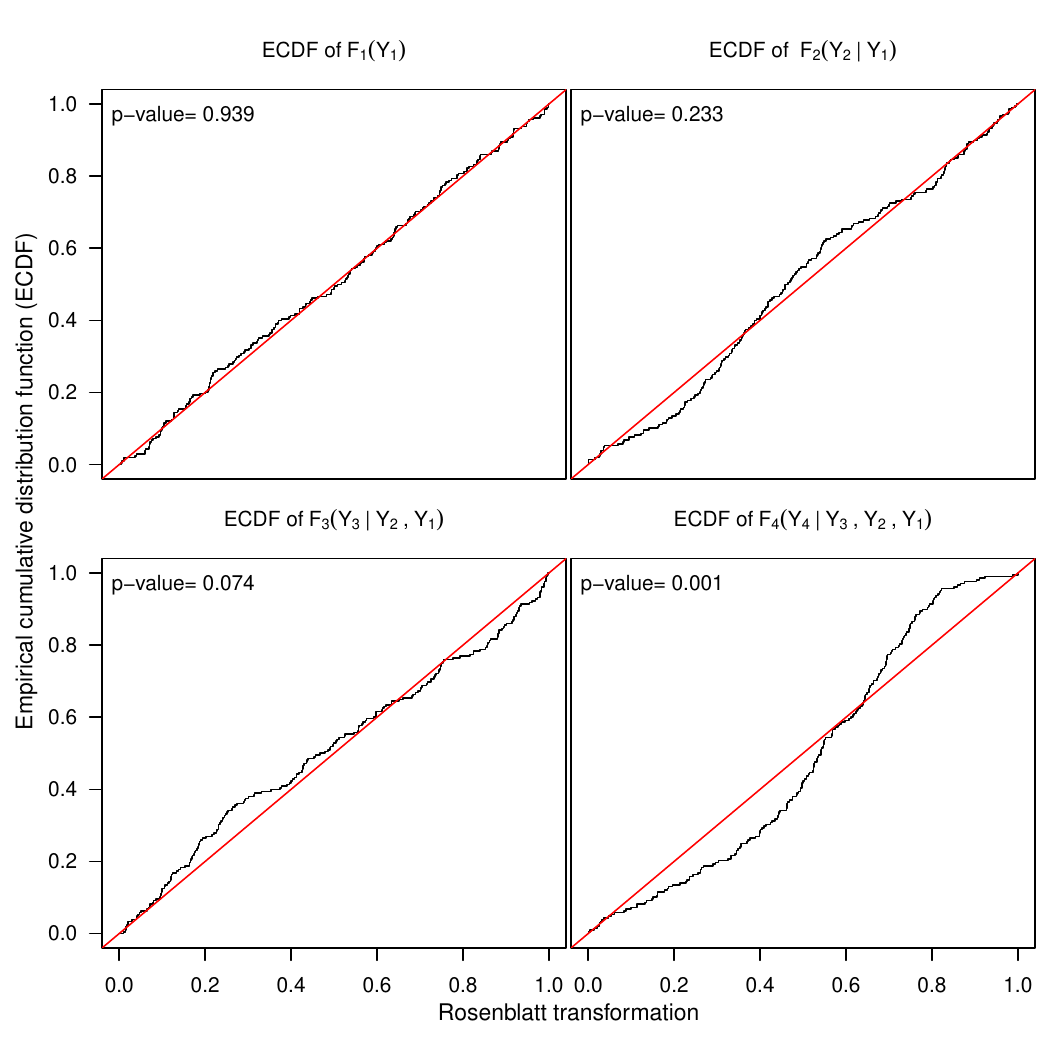} 
	\caption{Model diagnostic plots for the multivariate transformation model with 
                location-scale marginals and global correlation matrix (lsTDA) in subjects with hepatocellular carcinoma, employing the Rosenblatt transformation. The black line represents the empirical cumulative distribution function (ECDF) of the marginal Rosenblatt transformed data, while the red line represents the theoretical CDF of a uniform distribution.}
	\label{fig:mod_gof1}
\end{figure}

Figure~\ref{fig:mod_gof0} and~\ref{fig:mod_gof1} show the empirical CDFs of each of the transformed conditional margins $\hat{F}_j(Y_j \mid Y_1, \dots, Y_{j-1})$ in comparison to the theoretical CDF of the uniform distribution for cases without and with HCC, respectively. Additionally, for each margin, a Kolmogorov test for testing uniformity is conducted and the corresponding $p$-values are reported on the figure.

\subsubsection{Monotonicity checks}
\begin{figure}
	\centering
	\includegraphics[width=\linewidth]{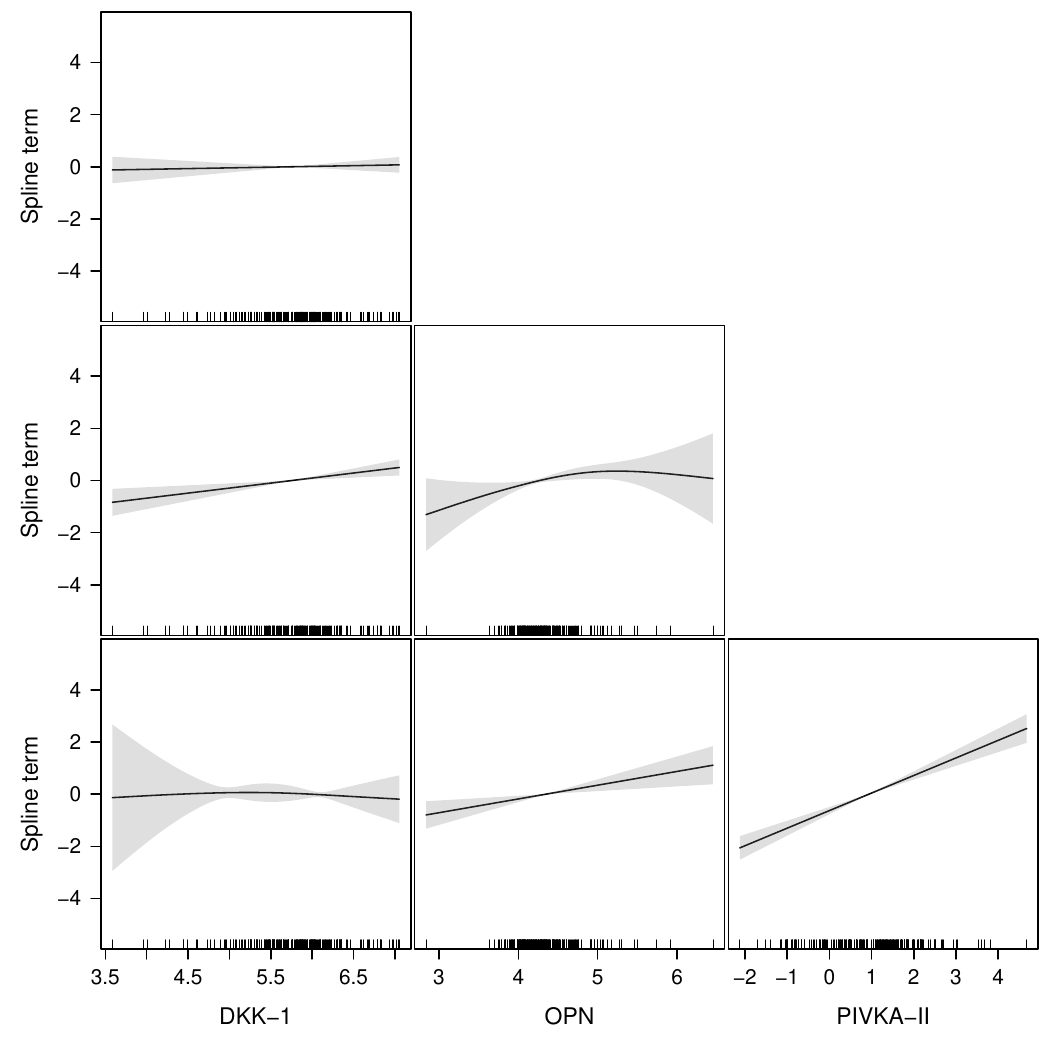} 
	\caption{Estimated transformation functions $g_{jk}$ for $2 \le k < j \le 4$ as a function of the conditioned biomarker values in subjects without hepatocellular carcinoma.}
	\label{fig:tramme0}
\end{figure}

\begin{figure}
	\centering
	\includegraphics[width=\linewidth]{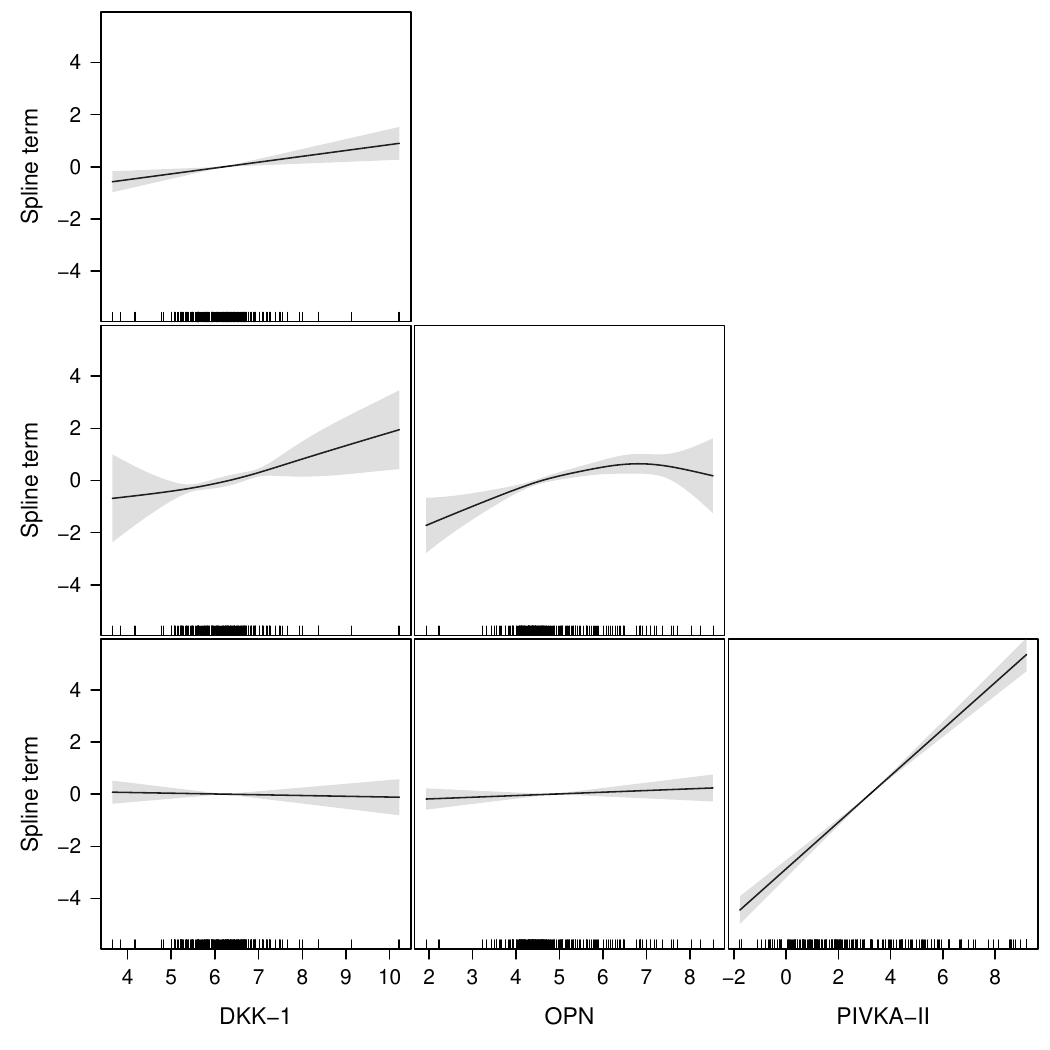} 
	\caption{Estimated transformation functions $g_{jk}$ for $2 \le k < j \le 4$ as a function of the conditioned biomarker values in subjects with hepatocellular carcinoma.}
	\label{fig:tramme1}
\end{figure}

Parametric copula models are restricted in capturing different types of
dependencies. Unreliable estimates may arise if any of the transformation functions
lacks monotonicity. To address this, we use thin plate splines in sequential univariate
additive transformation models \citep{tramme} to estimate the transformations $g_{jk}$
(which, under a Gaussian copula model, are $g_{jk} = \mLambda_{jk}
\h_{j}(y_{j})$, see above) in the model
\begin{align*}
	\hat{F}_j(y_j \mid \yvec_{j-1}) &= \Phi\left(\sum_{k = 1}^j g_{jk}(y_{j})\right)  \quad \text{for } \; j = 2,\dots,J.
\end{align*}
Figure~\ref{fig:tramme0} and~\ref{fig:tramme1} plot the estimated
transformation functions for $2 \le k < j \le 4$ as a function of the conditioned biomarker values
in subjects without and with hepatocellular carcinoma, respectively.  We use
these plots to check if there are any violations of monotonicity, since the
sum of monotone functions is also monotone.

In most cases, this assumption appears to hold, particularly when
considering the adequate estimation of spline terms with sufficient data. 
However, the regression of the OPN biomarker suggests that extrapolating the
model to larger biomarker values may result in unreliable estimates.  A
comprehensive validation study is essential before implementing the model
for clinical use.

\clearpage

\section{Additional results}
\label{sec:add}
\begin{figure}[h!]
	\centering
	\includegraphics[width=\linewidth]{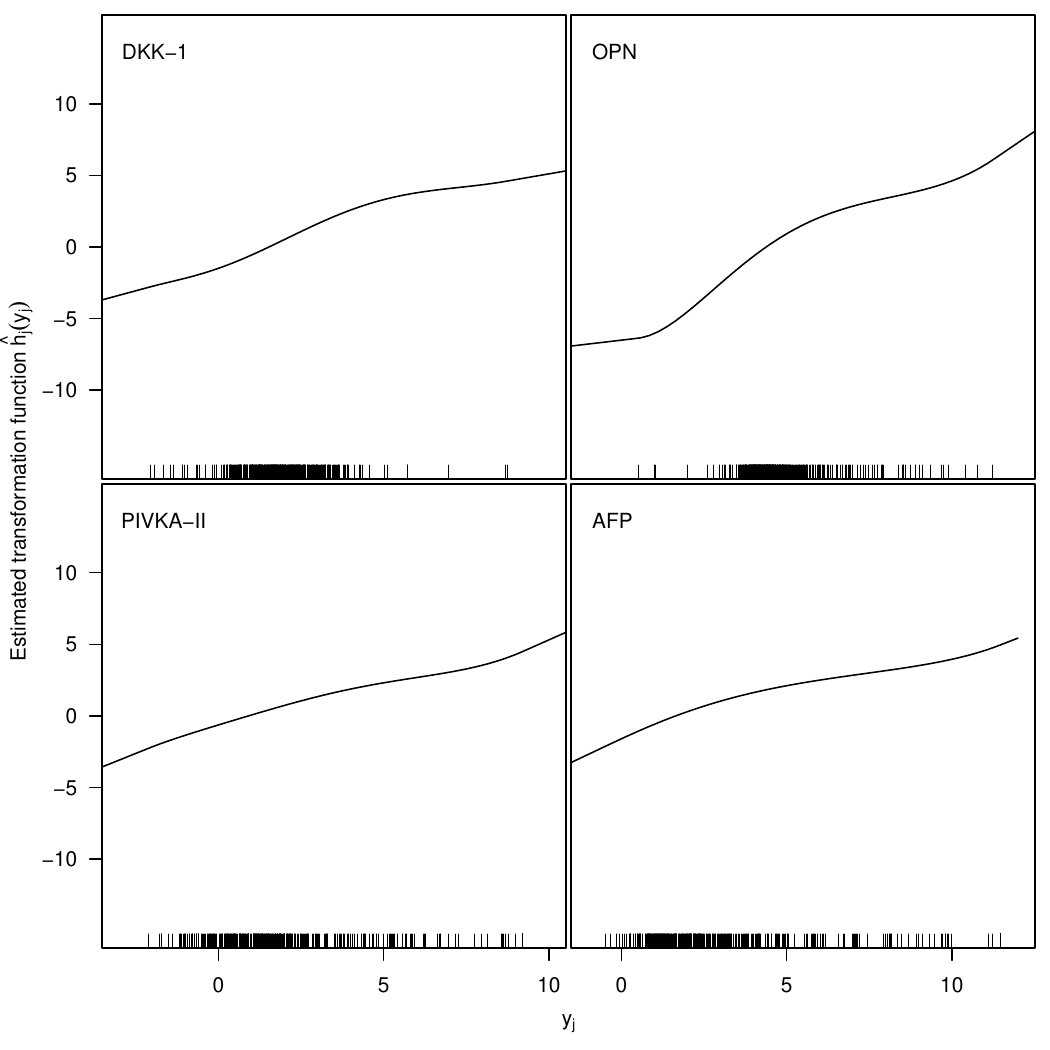} 
	\caption{Estimated transformation functions for the multivariate transformation model with 
                location-scale marginals and global correlation matrix (lsTDA) of hepatocellular carcinoma biomarkers.}
	\label{fig:trafo}
\end{figure}

\begin{figure}
	\centering
	\includegraphics[width=\linewidth]{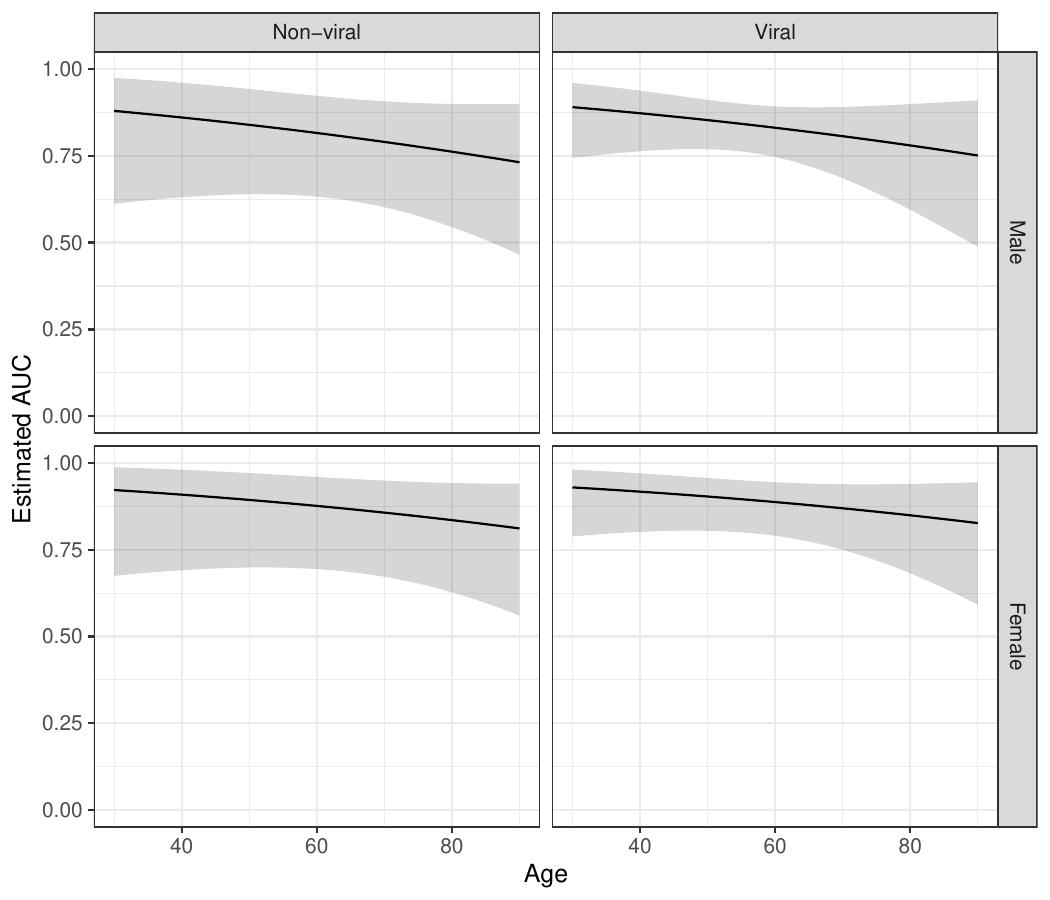} 
	\caption{Estimated covariate-dependent AUCs for the conditional class
	probabilities from the random forest, segmented by age and etiological groups, distinguishing between viral causes (HBV, HCV) and other etiologies such as alcohol-related or cryptogenic factors.}
	\label{fig:auc_covRF}
\end{figure}

Figure~\ref{fig:trafo} depicts the estimated marginal transformation
functions $\h_1, \dots, \h_4$ from the lsTDA model of HCC biomarkers.  These transformation
functions can be used in combination with the coefficients from
Table~\ref{tab:coefs} to calculate log-likelihood ratio score for a new subject. 
Figure~\ref{fig:auc_covRF} shows the estimated covariate-dependent AUCs
segmented by age and etiological groups for HCC biomarkers using a random
forest.  These results are presented for comparative analysis against our
method, as depicted in Figure~\ref{fig:auc_cov}, serving as a sensitivity
check between the two methods.
 \end{appendix}

\clearpage

\end{document}